\documentclass[11pt,notitlepage]{article}
\usepackage{amsmath}
\usepackage{epsfig}
\usepackage{graphics}
\usepackage[text={6.2in,9.4in},centering]{geometry}
\usepackage{color}
\usepackage{enumitem}
\usepackage{amssymb}

 \title{\bf A New Paradigm for Hadronic Parity Nonconservation and
 its Experimental Implications}
\author{Susan Gardner$^{a,d}$, W. C. Haxton$^{b,d}$, and Barry R. Holstein$^{c,d}$\\
\\
$^a$Department of Physics and Astronomy,\\
University of Kentucky,
Lexington, KY 40506\\
\\
$^b$Department of Physics, University of California, and\\
Lawrence Berkeley National Laboratory,
Berkeley, CA 94720\\
\\
$^c$Department of Physics-LGRT\\
University of Massachusetts,
Amherst, MA 01003\\
\\
$^d$Kavli Institute for Theoretical Physics\\
University of California,
Santa Barbara, CA  93016}

\begin{document}

\maketitle

\begin{abstract}
For decades the primary experimental goal in studies of hadronic parity nonconservation (PNC) has been the
isolation of the isovector weak nucleon-nucleon interaction, expected to be dominated by long-range pion exchange and
enhanced by the neutral
current.  In meson-exchange descriptions this interaction together with an isoscalar interaction generated by $\rho$
and $\omega$ exchange dominate most observables.  Consequently these two amplitudes have been used
to compare and check the consistency of the field's experiments.  Yet to date, despite sensitive searches like
that performed with $^{18}$F,  no evidence for isovector hadronic PNC has been found.
Here we argue, based on recent large-$N_c$ treatments and new global analyses,
that the emphasis on isovector hadronic PNC was misplaced.  Large-$N_c$ provides an alternative
and theoretically better motivated simplification of
effective field theories (EFTs) of hadronic PNC, separating the five low-energy constants (LECs) into two of leading order (LO),
and three others that are N$^2$LO.  This scheme pivots the isospin coordinates we have traditionally used,
placing one dominant axis in the isoscalar plane, and a second along the isotensor direction.
We show that this large-$N_c$ LEC hierarchy accurately describes all existing data on hadronic PNC.  In particular,
the null result found in $^{18}$F reflects its dependence on an N$^2$LO observable.
We discuss opportunities to further test the predicted large-$N_c$ hierarchy of LECs, illustrating the kind of analyses
experimentalists can use to better constrain the LO theory and to determine the size of N$^2$LO corrections.   This formalism --
combined with a new wave of experiments that will be performed at the SNS cold neutron beam line and 
the recent demonstration that lattice QCD can now be applied to PNC NN scattering -- could lead to rapid progress
in the next five years.  We discuss the impact of anticipated new results, including NPDGamma and a lattice QCD calculation
of isotensor PNC.  We also describe future experiments that can yield more precise 
values of the LO LECs and help isolate the N$^2$LO $\sim$ 10\% corrections.
\end{abstract}

\clearpage

\section{Introduction}

For more than two decades the field of hadronic parity nonconservation (PNC) has struggled: the theoretical picture has been
muddled, and very few new experiments have been done to help clarify matters.  This situation is about to change.
The creation of a high-intensity cold neutron beamline at the Spallation Neutron Source (SNS) as well as upgrades at NIST
will enable a new generation of experiments.  Results from the first experiment performed at the SNS,
the NPDGamma Collaboration's measurement of the
gamma ray asymmetry $A_\gamma(\vec{n}p\rightarrow d\gamma)$, should be announced soon~\cite{Crawford16,Fry:2016esm,Fry16}.  Furthermore, the feasibility of a new kind of ``measurement" -- evaluation of PNC NN scattering amplitudes from lattice QCD --
has been demonstrated \cite{Sci83}.

While such progress was being made, other developments have brought more clarity to the field,
changing some of our prejudices, and providing a new context for interpreting future results.
First, it was discovered that the agreement between past experiments was better than had been recognized \cite{Haxton:2013aca}:  most of the
tension that had existed was induced by flaws in global analyses of PNC.
Second, a new proposal emerged for organizing the low-energy
constants (LECs) that characterize hadronic PNC, based on large-$N_c$ QCD  \cite{Phillips:2014kna,Schindler:2015nga}.  We show here
that this proposal accounts simply for all existing data -- while illustrating that some of our past prejudices about
hadronic PNC patterns were not justified.    We describe these developments, and discuss their implications
for past and anticipated experiments, performing a new global
experimental analysis that exploits the large-$N_c$ LEC hierarchy.

NPDGamma is the latest effort to address one of the field's primary goals, measurement of PNC in the isovector
nucleon-nucleon (NN) interaction.  In meson-exchange descriptions this interaction is dominated by long-range
pion-exchange, with one $\pi$NN vertex governed by the weak interaction and the other by the strong.  The weak $\pi$NN vertex $h_\pi^1$ 
has been of particular interest because of the expectation it could help separate the roles of W- and Z-exchange in hadronic PNC,
associated with the four-currents $J_W$ and $J_Z$ that appear in the low-energy current-current Hamiltonian \cite{Donoghue:1992dd,Maiani:2011zz}
\begin{equation}
\mathcal{H} ={G_F \over \sqrt{2}} \left[ J_W^\dagger J_W + J_Z^\dagger J_Z \right] .
\end{equation}
The charged current can be decomposed into two components, $J_W=\cos{\theta_c} J_W^0 + \sin{\theta_c} J_W^1$, where $\theta_c$
is the Cabibbo angle.   The current $J_W^0$ drives the $u \rightarrow d$ transition and carries isospin and strangeness
$\Delta I=1$, $\Delta S = 0$, while $J_W^1$ drives the $u \rightarrow s$ transition and carries $\Delta I ={1 \over 2}$,
$\Delta S=-1$.  The neutral current also has two components, $J_Z= J_Z^0+J_Z^1$, transforming as $\Delta I=0$, $\Delta S=0$ and
$\Delta I=1$, $\Delta S=0$, respectively.
The current-current weak $\Delta S=0$ NN interaction is then
\begin{equation}
\mathcal{H}^{\Delta S =0} ={G_F \over \sqrt{2}} \left[ \cos^2{\theta_c} J_W^{0 \, \dagger} J_W^0 + \sin^2{\theta_c} J_W^{1 \, \dagger} J_W^1 + J_Z^{0 \, \dagger} J_Z^0 + J_Z^{1 \, \dagger} J_Z^1 +J_Z^{0 \, \dagger} J_Z^1 + J_Z^{1 \, \dagger} J_Z^0  \right] .
\end{equation}
The importance of isospin to this Lagrangian is clear: the symmetric product of two $\Delta I=1$ $J_W^0$ currents
transforms as $\Delta I=0$ and 2,  while the symmetric product of two $\Delta I={1 \over 2}$ $J_W^1$ currents transforms as
$\Delta I=1$.  Consequently the charged-current contribution to the $\Delta I=1$ weak NN interaction is suppressed by $\tan^2{\theta_c} \sim 0.04$,
relative to the charged-current contributions
to the $\Delta I=0$ and 2 interactions.  Based on this argument it was concluded that the $\Delta I=1$ component of the NN
interaction would be dominated by the neutral current, which is of particular interest because it cannot be studied in
strangeness-changing interactions due to the absence of tree-level flavor-changing neutral currents.
The expectation of an important
neutral current contribution to the $\Delta I=1$
PNC NN interaction has influenced PNC analyses since the 1980s.

The opportunity to take advantage of nuclei as laboratories to test an otherwise unconstrained standard-model interaction has
motivated both experiment and theory.  By the mid-1980s several advances had occurred:
\begin{enumerate}
\item Credible meson-exchange models of HPNC had been developed that established ``best values" and reasonable
ranges for hadronic PNC weak couplings (e.g., the DDH couplings of Desplanques, Donoghue, and Holstein \cite{Desplanques:1979hn}),
including $h_\pi^1$;
\item A series of successful experiments measuring the circular polarization of the gamma ray
from the decay of the 1.081 MeV J$^\mathrm{P}$T=$0^-0$ level in $^{18}$F had been
performed, testing the T=1 mixing of this state
with the nearby $0^+1$ level at 1.042 MeV \cite{Barnes:1978sq, Biz80, Ahrens:1982,Bini:1985zz,Page:1987ak}, but finding no
signal at the $ \sim 10^{-4}$ level; and
\item A method to extract a limit on $h_\pi^1$ from these measurements was developed, exploiting a relationship between hadronic PNC and axial-charge $\beta$-decay, largely eliminating any dependence on the
choice of nuclear wave functions \cite{Haxton:1981sf,Bennett80,Adelberger:1983zz}.
Several high-quality measurements
 of the $\beta$ decay rate were made \cite{Adelberger:1983zz}.
\end{enumerate}
This led to a surprising conclusion: the $^{18}$F result, by itself or when combined with other measurements
in a general hadronic PNC analysis \cite{Adelberger:1985ik}, established an upper bound on $h_\pi^1$,
relative to competing isoscalar amplitudes,
about a factor of six below the DDH best value.  The isoscalar amplitudes are somewhat
stronger than expected, and the isovector considerably weaker.  The analogy between this
result and the
$\Delta I ={1 \over 2}$ rule in strangeness-changing decays -- where a similar anomaly in the ratio
of $\Delta I={3 \over 2}$ to $\Delta I={1 \over 2}$ amplitudes is found -- was immediately noted \cite{Adelberger:1985ik}.
These results show that the neutral current, once
embedded in the strongly interacting environment of the nucleon, produces an effective coupling to the nucleon
considerably weaker than that expected from the elementary coupling to the quarks.

These results helped motivate the NPDGamma effort.
The Collaboration has worked toward its goal of measuring $h_\pi^1$ since approximately 2000,
beginning with an earlier version of the experiment at LANSCE \cite{Gericke:2011zz},
then moving to the SNS to exploit the high-intensity cold neutron beamline
that became available there \cite{Crawford16,Fry:2016esm,Fry16}.  
The goal has been to reach a sensitivity corresponding to $A_\gamma \sim 10^{-8}$ \cite{Page05}, verifying the $^{18}$F
result and possibly pushing beyond, to a detection of $h_\pi^1$.  Experimental issues at LANSCE, including a weaker
than anticipated neutron flux, led to a final result of
$A_\gamma = (-1.2 \pm 2.1 \mathrm{(stat)} \pm 0.2 \mathrm{(sys)}) \times 10^{-7}$ \cite{Gericke:2011zz}.  
The effort to reach the $10^{-8}$
level was renewed on the high-intensity SNS beamline \cite{Crawford}.  The  DDH best-value prediction for 
$A_\gamma$ is $\sim 5 \times 10^{-8}$.

The purpose of this paper is to re-examine past experiments and consider anticipated new
results, including those from NPDGamma, in the context of a new paradigm for hadronic PNC
that arises when effective field theory (EFT) is combined with large-$N_c$ instructions 
for treating the EFT's LECs.  In the context of this
scheme, not only $^{18}$F but the entire set of hadronic PNC observations
conform to a simple pattern. The purpose of this article is describe recent developments
that have led to this satisfying conclusion.  These include
\begin{enumerate}
\item Modified constraints \cite{Haxton:2013aca} extracted from the longitudinal asymmetry $A_L(\vec{\mathrm{p}}\mathrm{p})$, at both
low and intermediate energies, obtained when certain inconsistencies in past global hadronic PNC analyses
were corrected.
\item An improved understanding of EFT approaches to hadronic PNC, ironing out some apparent differences 
between EFT descriptions \cite{Zhu:2004vw,Girlanda:2008ts,Phillips:2008hn} as well as establishing their effective equivalence to the traditional meson exchange treatment,
when the latter is restricted to low energies.  The EFT approach leads to a description quite similar to Danilov's \cite{Danilov:1965,Danilov:1971fh,Danilov:1972}
early analysis in terms of S-P amplitudes, involving five degrees of freedom (or perhaps six
if nuclear systems like $^{18}$F are used, where the pion's range is important).
\item As we lack the data needed to perform a five- or six-dimensional analysis, the field has long employed
simpler analyses with a smaller number of ``most important" couplings.
In the DDH meson-exchange model, $h_\pi^1$ and a corresponding vector-meson isoscalar
coupling have been employed as the leading terms.  However, as we will discuss, this choice 
is both in conflict with experiment and lacking in theoretical justification.
Recently an alternative organizational
scheme has been proposed, based on the large $N_c$ expansion, that appears to be
compatible with all we have learned about hadronic PNC to date.  This expansion identifies
two alternative leading couplings, and 
predicts that subleading corrections to this scheme will enter at the relative order $1/N_c^2 \sim$ 10\%.
\item This combined EFT/large-$N_c$ approach leads to a lowest-order (LO) two-dimensional (2D) characterization of hadronic PNC
that uses the isotensor${}^1S_0-{}^3P_0$ amplitude together with a specific combination of the isoscalar
${}^3S_1-{}^1P_1$ and ${}^1S_0-{}^3P_0$ amplitudes.   We derive the LO  LECs
from current experiments, and show the general concordance between PNC
observations and the coupling hierarchy predicted by the large-$N_c$ scheme.
\item We describe further opportunities to further test this picture, through new experiments
and by using lattice QCD.
\item  The $^{18}$F and NPDGamma experiments probe observables
that are blind to the LO couplings, and thus arise only through N$^2$LO corrections.
Thus the absence of a $^{18}$F signal is not in conflict with theory, but is in fact consistent
with large-$N_c$ predictions.  We show that by combining  the $^{18}$F result with the anticipated
NPDGamma measurement, important constraints can be placed on N$^2$LO couplings.
\end{enumerate}
This paper reviews the first three developments above, then addresses the last three points
by performing a new global analysis using the EFT/large-$N_c$ formalism.

\section{Background}
The violation of parity invariance, suggested by Lee and Yang in 1956~\cite{Lee:1956qn}, was discovered experimentally in 1957 by Wu et al. via measurement of the
$\boldsymbol{J}\cdot\boldsymbol{p}_e$ correlation parameter in the beta decay of oriented $^{60}$Co~\cite{Wu:1957my}.  It was immediately recognized that there should exist a corresponding PNC component of the NN interaction.
The first experimental search for PNC was conducted by Tanner, who in 1957 sought evidence in the ${}^{19}{\rm F}(p,\alpha)^{16}{\rm O}$ reaction~\cite{Tanner:1957zz}.  Although the sensitivity of this measurement was not sufficient to observe a PNC signal, it was the first of a series of such experiments that have continued to the present time, to detect subtle
weak interaction effects in systems with strong and electromagnetic interactions. A summary can be found in various review articles~\cite{Adelberger:1985ik,Haeberli:1995uz,RamseyMusolf:2006dz,Haxton:2013aca}.
PNC has been observed in many hadronic systems, usually in the form of some pseudoscalar such as an
asymmetry or a circular polarization, including cases where rather spectacular enhancements
of the signal arise.  Examples include the $2\%$ photon circular polarization in the electromagnetic decay of an isomer of ${}^{180}{\rm Hf}$~\cite{Krane:1971zz}
\begin{equation}
A_\gamma({}^{180}{\rm Hf}^*\rightarrow{}^{180}{\rm Hf}+\vec{\gamma})=-(1.66\pm 0.18)\times 10^{-2},
\end{equation}
and the nearly $10\%$ longitudinal analyzing power for the scattering of polarized neutrons from $^{139}{\rm La}$~\cite{Yuan:1991zz}
 \begin{equation}
A_h(\vec{n}+{}^{139}{\rm La})=(9.55\pm 0.35)\times 10^{-2}.
\end{equation}
These amplifications originate from chance nuclear level near degeneracies: states of the same spin but opposite parity can mix through the hadronic PNC interaction, generating a parity admixture inversely proportional to the
energy splitting between the levels.  Indeed, as the natural scale of hadronic PV effects is $\sim G_Fm_\pi^2\sim 10^{-7}$~\cite{Adelberger:1985ik}, it is apparent that such enhancements can be many orders of
magnitude.

With the development of theoretical frameworks for understanding hadronic PNC quantitatively, the
field's attention switched from simply finding examples of PNC, to identifying systems where PNC could be
both measured and reliably interpreted in terms of the underlying PNC NN interaction.  In an ideal
world such measurements would be carried out by studying the simple NN systems -- pp, np, and pn --
in the allowed threshold partial-wave channels.  As only one such measurement has been done at high
precision, the field turned to few-nucleon systems, where techniques exist for solving the
Schr\"{o}dinger equation with realistic strong potentials.   Certain selected light nuclei also
provided reliable hadronic PNC constraints.  Notable among these are $^{18}$F and $^{19}$F, where parity
doublets enhance the experimental signal, and axial-charge beta decay measurements can remove almost
all of the usual nuclear structure uncertainties.

\subsection{The DDH potential}
Helping to drive these developments were theoretical descriptions of hadronic PNC defining
what needed to be measured, and providing benchmark estimates of potential hadronic PNC signal sizes.
The ``standard" formalism for hadronic PNC experimental analysis became the meson exchange model of Desplanques, Donoghue, and Holstein (DDH)~\cite{Desplanques:1979hn}, which expresses the hadronic PNC potential in terms of the usual parity-conserving strong interaction
meson-nucleon couplings defined by\small
\begin{eqnarray}
{\cal H}_{\rm st}&=& ig_{\pi NN}\bar{N}\gamma_5\boldsymbol{\tau}\cdot\boldsymbol{\pi} N
+g_\rho\bar{N}\left(\gamma_\mu+i{\chi_\rho\over
2m_N}\sigma_{\mu\nu}k^\nu\right)
\boldsymbol{\tau}\cdot\boldsymbol{\rho}^\mu N\nonumber\\
&&+g_\omega\bar{N}\left(\gamma_\mu
                        +i{\chi_\omega\over 2m_N}\sigma_{\mu\nu}k^\nu
\right)\omega^\mu N, \label{eq:pch}
\end{eqnarray}\normalsize
where $m_N$ is the nucleon mass.  The numerical values assigned to the various couplings by DDH are $g^2_{\pi NN}/4 \pi = 14.4$,
$g_\rho^2/4 \pi={1 \over 9} {g_\omega^2}/4 \pi =0.62 $, $\chi_\rho=\kappa_p-\kappa_n=3.70$, and $\chi_\omega=\kappa_p+\kappa_n=-0.12$.
DDH employed the phenomenological PNC weak meson-nucleon Hamiltonian
\begin{eqnarray}
&&{\cal H}_{\rm wk} =i{h_\pi^1\over
\sqrt{2}}\bar{N}(\boldsymbol{\tau}\times\boldsymbol{\pi})_zN\nonumber\\
&+&\bar{N}\left(h_\rho^0\boldsymbol{\tau}\cdot\boldsymbol{\rho}^\mu +h_\rho^1\rho_z^\mu
+{h_\rho^2\over 2\sqrt{6}}(3\tau_z\rho_z^\mu
-\boldsymbol{\tau}\cdot\boldsymbol{\rho}^\mu)\right)
\gamma_\mu\gamma_5N\nonumber\\
&+&\bar{N}
\left(h_\omega^0\omega^\mu+h_\omega^1\tau_z\omega^\mu\right)\gamma_\mu\gamma_5N
-h_\rho^{'1}\bar{N}(\boldsymbol{\tau}\times\boldsymbol{\rho}^\mu)_z{\sigma_{\mu\nu}k^\nu\over
2m_N} \gamma_5N.
\label{eq:weak}
\end{eqnarray}\normalsize
When Eqs. (\ref{eq:pch}) and (\ref{eq:weak}) are combined and a nonrelativistic reduction is
made, a PNC potential is obtained with a specific spin and isospin structure, together with specific radial forms,
governed by the meson masses.   The coefficients of this potential are bilinears in the weak and
strong couplings -- these products are the parameters that can be extracted.

Provided a consistent set of strong couplings is employed in global analyses,
such a program would determine values for the weak couplings $h_\pi^1,h_\rho^{0,1,2},h_\rho^{1'},h_\omega^{1,2}$.
In their original work DDH provided theoretical estimates for these parameters --
while emphasizing their {\it very} large uncertainties.  DDH provided a best guess
for each parameter (called by these authors the ``best value"), as well as a ``reasonable range"
to provide a measure of the uncertainty, as shown in Table 1.
\begin{table}
\begin{center}
\caption{The weak meson-nucleon couplings as estimated in Refs.
\cite{Desplanques:1979hn,Dubovik:1986pj,Feldman:1991tj}. All numbers are quoted in units of $10^{-7}$.  Following
the original treatment by DDH, $h_\rho^{\prime \, 1}$ has been set to zero:  it can be shown at low energies 
that this coupling is redundant \cite{Haxton:2013aca}.}
\begin{tabular}{|c|c|c|c|c|}
\hline \quad   & DDH~\cite{Desplanques:1979hn} & DDH~\cite{Desplanques:1979hn} & DZ~\cite{Dubovik:1986pj} &
FCDH~\cite{Feldman:1991tj}\\
Coupling & Reasonable Range & ``Best" Value &  &  \\ \hline
$h_\pi^1$ & $0\rightarrow 11$ &+4.6&+1.1&+2.7\\
$h_\rho^0$& $11\rightarrow -31$&$-11$&$-8.4$&$-3.8$\\
$h_\rho^1$& $-0.4\rightarrow 0$& $-0.2$&+0.4&$-0.4$\\
$h_\rho^2$& $-7.6\rightarrow -11$&$-9.5$&$-6.8$&$-6.8$\\
$h_\omega^0$&$5.7\rightarrow -10.3$&$-1.9$&$-3.8$&$-5.0$\\
$h_\omega^1$&$-1.9\rightarrow -0.8$&$-1.2$&$-2.3$&$-2.3$\\ \hline
\end{tabular}
\end{center}
\label{tab0}
\end{table}

Despite a great deal of experimental and theoretical work, most of this uncertainty remains today.
The main obstacle to reducing the reasonable ranges has been a lack of reliable experimental constraints.  Critical examinations of the available data
have led to the conclusion that only four experiments place important constraints on
hadronic PNC -- the analyzing power for the scattering of longitudinally polarized
protons off protons, the circular polarization of the
gamma ray emitted from the $0^-0$ excited state in $^{18}$F, the analyzing power for the scattering of longitudinally polarized
protons off $^4$He, and the gamma decay asymmetry from the decay of the polarized ${1 \over 2}^-{1 \over 2}$
state in $^{19}$F.  The last two measurements both involve odd-proton systems, and
consequently yield almost the same constraint, while the $^{18}$F result is only an upper bound
(though a very significant one, given the DDH best-value estimate of the best value for $h_\pi^1$).

To improve on the current situation, one needs either several new, interpretable experimental results, or a strategy
that reduces the number of theoretical variables that must be determined.  Such a strategy has recently been proposed,
based in large-$N_c$ QCD.  As this approach is most naturally described in terms of effective field theory treatments
of hadronic PNC, we first describe that formalism and its relationship to the DDH potential.

\subsection{The EFT Picture}

While the DDH meson-exchange approach clearly contains some model dependence, it has stood as
the standard language for analyzing low energy PNC experiments for nearly four decades.
In recent years, however, an alternative to the DDH potential has been developed, based on pionless effective
field theory.  Pionless EFT provides a model-independent formalism for describing experiments performed at momentum scales well
below the pion mass, where the pion interaction becomes local.  Most applications to 
PNC scattering satisfy this condition \cite{Schindler:2013yua}, at least
with respect to the external momenta of the scattered particles.  (If a nuclear bound state is involved, however,
the nuclear Fermi momentum can also play a role.)

This approach was introduced in studies of hadronic PNC by Zhu et al.~\cite{Zhu:2004vw}, though the roots of this kind of
analysis reach back to Danilov's partial wave analysis \cite{Danilov:1965,Danilov:1971fh} and to the use of contact
potentials by Desplanques and
Missimer \cite{Desplanques:1976mt}.  The Zhu et al. formulation contained redundant terms that were
later identified and eliminated by Girlanda \cite{Girlanda:2008ts}.  The EFT method was also developed in the work of
Phillips, Schindler, and Springer \cite{Phillips:2008hn}.  In pionless EFT the
NN interaction is represented by a small number of empirically determined contact terms. In the parity-conserving case, for example, there are only two, representing scattering lengths in the $^1S_0$ and $^3S_1$ channels.
In the parity-violating case, however, there exist {\it five} low-energy S-P channels, and consequently
five associated LECs.

While the DDH and EFT approaches appear to be quite distinct, in fact they are operationally equivalent at
the very low energies where pionless EFT is valid.  This point was recently made by constructing
an effective contact interaction that maps onto Danilov's partial wave analysis~\cite{Haxton:2013aca}
\begin{eqnarray}
V^{PNC}_{LO}(\boldsymbol{r})&=&\Lambda_0^{^1S_0-^3P_0}\left({1\over i}{\overleftrightarrow{\boldsymbol{\nabla}}_A\over 2m_N}{\delta^3(\boldsymbol{r})\over m_\rho^2}\cdot(\boldsymbol{\sigma}_1-\boldsymbol{\sigma}_2)
-{1\over i}{\overleftrightarrow{\boldsymbol{\nabla}}_S\over 2m_N}{\delta^3(\boldsymbol{r})\over m_\rho^2}\cdot i(\boldsymbol{\sigma}_1\times\boldsymbol{\sigma}_2)\right)\nonumber\\
&+&\Lambda_0^{^3S_1-^1P_1}\left({1\over i}{\overleftrightarrow{\boldsymbol{\nabla}}_A\over 2m_N}{\delta^3(\boldsymbol{r})\over m_\rho^2}\cdot(\boldsymbol{\sigma}_1-\boldsymbol{\sigma}_2)
+{1\over i}{\overleftrightarrow{\boldsymbol{\nabla}}_S\over 2m_N}{\delta^3(\boldsymbol{r})\over m_\rho^2}\cdot i(\boldsymbol{\sigma}_1\times\boldsymbol{\sigma}_2)\right)\nonumber\\
&+&\Lambda_1^{^1S_0-^3P_0}\left({1\over i}{\overleftrightarrow{\boldsymbol{\nabla}}_A\over 2m_N}{\delta^3(\boldsymbol{r})\over m_\rho^2}\cdot(\boldsymbol{\sigma}_1-\boldsymbol{\sigma}_2)(\tau_{1\, z}+\tau_{2 \, z})\right)\nonumber\\
&+&\Lambda_1^{^3S_1-^3P_1}\left({1\over i}{\overleftrightarrow{\boldsymbol{\nabla}}_A\over 2m_N}{\delta^3(\boldsymbol{r})\over m_\rho^2}\cdot(\boldsymbol{\sigma}_1+\boldsymbol{\sigma}_2)(\tau_{1 \, z}-\tau_{2 \, z})\right)\nonumber\\
&+&\Lambda_2^{^1S_0-^3P_0}\left({1\over i}{\overleftrightarrow{\boldsymbol{\nabla}}_A\over 2m_N}{\delta^3(\boldsymbol{r})\over m_\rho^2}\cdot(\boldsymbol{\sigma}_1-\boldsymbol{\sigma}_2)
(\boldsymbol{\tau}_1\otimes\boldsymbol{\tau}_2)_{20}\right),
\label{eq:kj}
\end{eqnarray}
where $(\boldsymbol{\tau}_1 \otimes \boldsymbol{\tau})_{20} \equiv (3 \tau_{1 \,z} \tau_{2 \,z}-\boldsymbol{\tau}_1 \cdot \boldsymbol{\tau}_2)/\sqrt{6}$.  The subscripts on the LECs denote the change in isospin $\Delta I$ induced by the associated operator,
while the superscripts indicate the specific PNC transition.  With these operator definitions
the various $\Lambda$s are dimensionless.
Of course, there must exist a matching to the low-energy form of the DDH potential, yielding~\cite{Haxton:2013aca}
\begin{eqnarray}
\Lambda_0^{^1S_0-^3P_0}&=&-g_\rho(2+\chi_\rho)h_\rho^0-g_\omega(2+\chi_\omega)h_\omega^0          \hspace{3.4cm}^\mathrm{DDH}\Lambda_0^{^1S_0-^3P_0} = 210\nonumber\\
\Lambda_0^{^3S_1-^1P_1}&=&-3g_\rho\chi_\rho h_\rho^0+g_\omega\chi_\omega h_\omega^0 \hspace{5.1cm} ^\mathrm{DDH}\Lambda_0^{^3S_1-^1P_1}=360\nonumber\\
\Lambda_1^{^1S_0-^3P_0}&=&-g_\rho(2+\chi_\rho)h_\rho^1-g_\omega(2+\chi_\omega )h_\omega^1 \hspace{3.35cm} ^\mathrm{DDH}\Lambda_1^{^1S_0-^3P_0}=21\nonumber\\
\Lambda_1^{^3S_1-^3P_1}&=&\textstyle{\sqrt{1\over 2}}\, g_{\pi NN}\left({m_\rho\over m_\pi}\right)^2h_\pi^1
+g_\rho(h_\rho^1-{h_\rho^1}')-g_\omega h_\omega^1 \hspace{1.4cm} ^\mathrm{DDH}\Lambda_1^{^3S_1-^3P_1}= 1340\nonumber\\
\Lambda_2^{^1S_0-^3P_0}&=&-g_\rho(2+\chi_\rho)h_\rho^2 \hspace{6.04cm} ^\mathrm{DDH}\Lambda_2^{^1S_0-^3P_0}=160\label{eq:mn}
\end{eqnarray}
where on the right the DDH predicted ``best values" have been employed, yielding values for the LECs (units of $10^{-7}$). Similarly, the EFT potentials of
Girlanda et al.~\cite{Girlanda:2008ts} and Zhu et al.~\cite{Zhu:2004vw} must also be equivalent to Eq. (\ref{eq:kj}).  The translation between the various formulations is given in the ``Rosetta stone" Table 2 in \cite{Haxton:2013aca}.

This comparison shows that the DDH potential is effectively equivalent to pionless EFT at the low energies
for which the latter is valid.   In this regime an S-P partial wave description is adequate, and five linear
combinations of the seven DDH weak couplings describe the physics.  The redundancy among these
parameters is broken when P-D interactions become important.  Then the meson masses also play
an explicit role, as higher partial wave channels allow one to detect the non-contact form of the radial
interaction.  One can think of the DDH interaction as an EFT that is married to a physically
motivated model, for the purpose extending the interaction's range of validity to higher momenta.

However, regardless of what formulation one uses, there remains a major problem:  five parameters are
needed to describe hadronic PNC in the low-momentum limit, but we do not have five reliable
experimental constraints.  Thus some simplification is needed, beyond that provided by EFT
or by a low-momentum reduction of the DDH potential.


\subsection{Experimental Constraints and 2D Reductions}
A standard display of experimental constraints on hadronic PNC was introduced in \cite{Adelberger:1985ik} and has
been in broad use ever since.  It employs two parameters, not five, and was derived on largely
empirical grounds -- an after-the-fact examination of how theoretical predictions of PNC observables
depend on the underlying weak couplings.  In light of subsequent discussions we will have of
the large-$N_c$ expansion, we sketch here how this standard display came about.

\begin{figure}[t]
\centerline{\includegraphics[width=150mm]{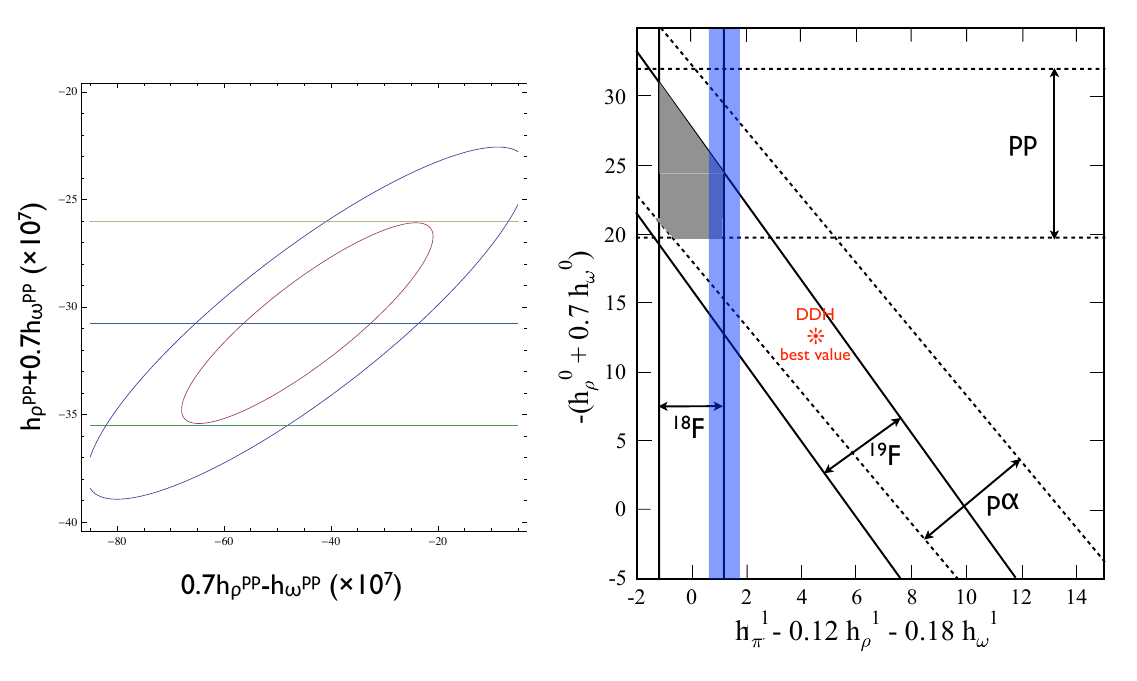}}
\caption{{Right panel: the standard plot of experimental constraints on the DDH parameters, including recent revisions introduced in \cite{Haxton:2013aca}. The experimental bands are $1\sigma$.} The blue band comes from a preliminary estimate of $h_\pi^1$, based on a lattice QCD calculation of a three-point function  \cite{SciWasem}, as discussed
in the text.
Left panel: constraints on $h_{\rho,\omega}^\mathrm{pp} = h_{\rho,\omega}^0+h_{\rho,\omega}^1+h_{\rho,\omega}^2/\sqrt{6}$ derived
from $A_L(\vec{\mathrm{p}}\mathrm{p})$ \cite{Haxton:2013aca,Carlson:2001ma}.
The ellipses are the 68\% and 90\% c.l. contours.}
\label{fig:standard}
\end{figure}

\begin{figure}[t]
\begin{center}
\includegraphics[width=120mm]{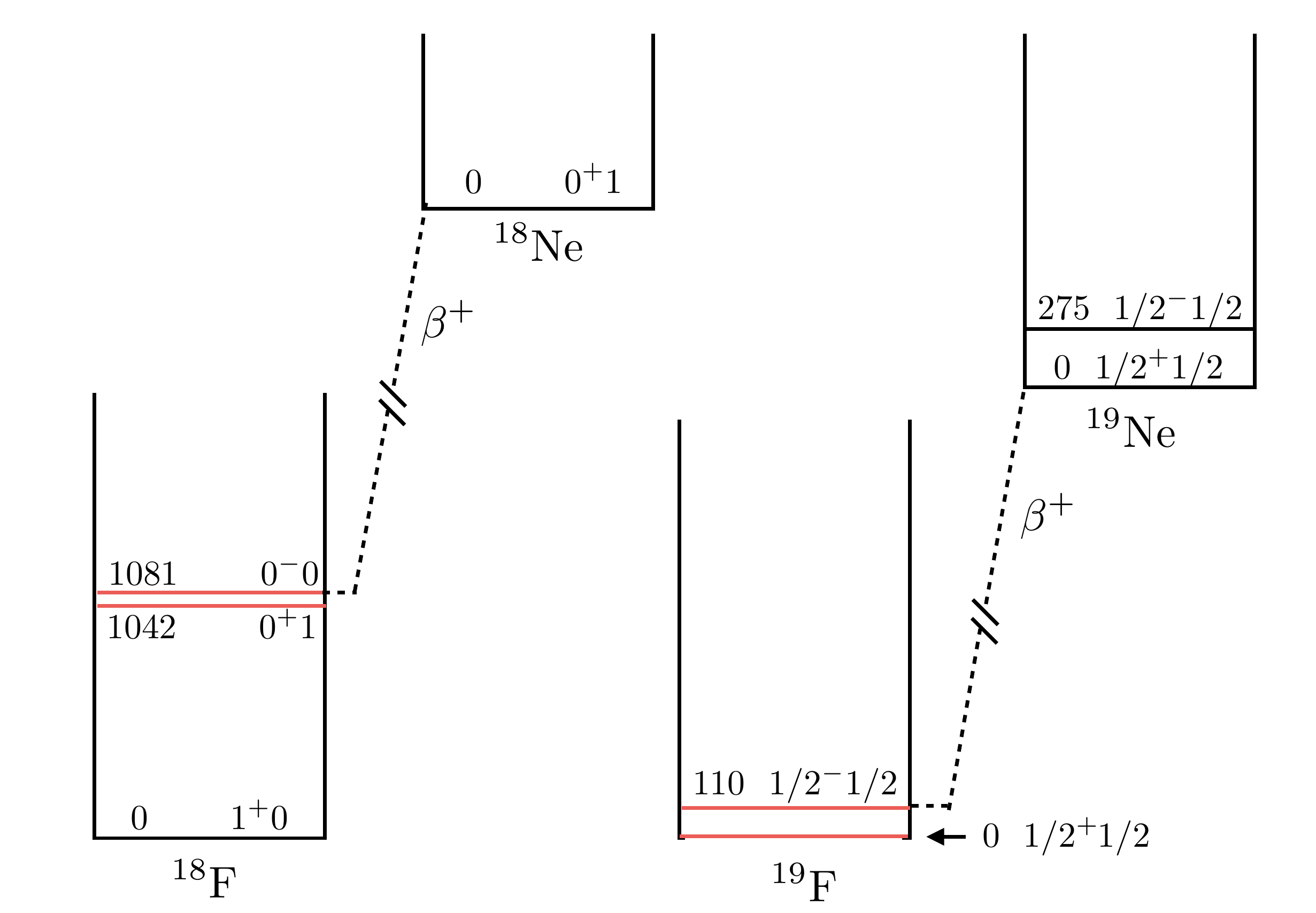}
\caption{The parity doublets in $^{18}$F and $^{19}$F are indicated in red.  The $\beta^+$ decays from $^{18}$Ne and
$^{19}$Ne connect the negative-parity members of the parity doublets to the isotopic analogs of the positive-parity members
(the ground states of $^{18}$Ne and $^{19}$Ne).  These axial-charge $\beta^+$ decays can be used to calibrate 
the strength of the PNC doublet mixing. Energies are in keV.}
\label{fig:F1819}
\end{center}
\end{figure}

The current version of this plot is shown in Fig. \ref{fig:standard}.   It includes constraints from
four experiments:
\begin{enumerate}
\item The longitudinal analyzing power in the scattering of polarized protons from an unpolarized proton target, for which measurements were 
performed at 13.6 MeV at Bonn, 15 MeV at LANL, 45 MeV at PSI, and 221 MeV at TRIUMF,
\begin{equation}
A_L(\vec{p}p)=\left\{\begin{array}{cc}
(-0.93\pm 0.20\pm 0.05)\times 10^{-7}& 13.6\,{\rm MeV}~\cite{Eversheim:1991tg}\\
(-1.7\pm 0.8)\times 10^{-7}&15\,{\rm MeV}~\cite{Nagle:1978vn}\\
(-1.57\pm 0.23)\times 10^{-7}&45\,{\rm MeV}~\cite{Balzer:1980dn,Balzer:1985au,Kistryn:1987tq}\\
(0.84\pm 0.34)\times 10^{-7}&221\,{\rm MeV}~\cite{Berdoz:2002sn,Berdoz:2001nu}
\end{array}\right. ~~.
\end{equation}
The first three experiments were done at relatively low energy, where a description in terms of
S-P amplitudes is a reasonable approximation.  Thus they constrain the partial wave coefficients
of Eq. (\ref{eq:kj}) in a straightforward way \cite{Haxton:2013aca,Carlson:2001ma},
\begin{equation}
\Lambda_0^{^1S_0-^3P_0}+\Lambda_1^{^1S_0-^3P_0}+{1\over \sqrt{6}}\Lambda_2^{^1S_0-^3P_0} =
419\pm 43~.
\end{equation}
The S-P LECs $\Lambda$ are given in units of $10^{-7}$.
The TRIUMF measurement, in contrast, must be treated in a formalism that includes
higher partial waves in the weak interaction, such as the DDH potential.
As indicated in the left panel of Fig. \ref{fig:standard}, the resulting constraints on weak
couplings can be expressed in terms of one combination related to S-P amplitudes, 
using the relations in Eq. (\ref{eq:mn}), and another associated with P-D amplitudes.
We will discuss this result in more detail below.

\item The longitudinal analyzing power for scattering 46 MeV
polarized protons on a $^4$He target was measured at PSI ~\cite{Lang:1985jv,Henneck:1982cc}
\begin{equation}
A_L(\vec{\rm p}\alpha)\big|_{46 \mathrm{MeV}}=-(3.3\pm 0.9)\times 10^{-7},
\end{equation}
placing the following constraint on the S-P LECs \cite{Lang:1985jv,Haxton:2013aca}
\begin{equation}
\Lambda_0^{^1S_0-^3P_0}+0.89\Lambda_1^{^1S_0-^3P_0}+0.75\Lambda_0^{^3S_1-^1P_1}+
0.32\Lambda_1^{^3S_1-^3P_1}=
930\pm 253~.
\end{equation}

\item The circular polarization of photons emitted in the decay of the 1.081 MeV $0^-0$ excited state of $^{18}$F to the $1^+0$ ground state is induced by PNC mixing of the $0^-$ state with the nearby
$0^+1$ state at 1.042 MeV (see Fig \ref{fig:F1819}).  Consequently this experiment selects out isovector hadronic PNC.
 Four independent experiments have yielded the limits
\begin{equation}
P_\gamma=\left\{\begin{array}{cc}
(-7\pm 20)\times 10^{-4}&{\rm Caltech/Seattle}~\cite{Barnes:1978sq}\\
(-10\pm 18)\times 10^{-4}&{\rm Mainz}~\cite{Ahrens:1982}\\
(3\pm 6)\times 10^{-4}& {\rm Florence}~\cite{Bini:1985zz}\\
(2\pm 6)\times 10^{-4}& {\rm Queens}~\cite{Page:1987ak}
\end{array}\right.~.
\end{equation}
These results lead to the constraint \cite{Haxton:1981sf,Haxton:2013aca}
\begin{equation}
|\Lambda_1^{^3S_1-^3P_1}+2.42 \Lambda_1^{^1S_0-^3P_0}|<340,\label{eq:gd}
\end{equation}
which implies a value of $h_\pi^1$ significantly below the DDH best value.

\item The gamma decay angular asymmetry for the transition from the polarized 110 keV ${1\over 2}^-$ excited state
in $^{19}$F to the ${1\over 2}^+$ ground state has been measured, testing the parity mixing of these
levels (see Fig. \ref{fig:F1819}).  The results
\begin{eqnarray}
A_\gamma=\left\{\begin{array}{cc}
(-8.5\pm 2.6)\times 10^{-5}&{\rm Seattle}~\cite{Adelberger:1983zz}\\
(-6.8\pm 1.8)\times 10^{-5}&{\rm Mainz}~\cite{Elsener:1987sx,Elsener:1984vp}
\end{array}\right.
\end{eqnarray}
constrain isoscalar and isovector hadronic PNC \cite{Adelberger:1983zz,Haxton:2013aca}
\begin{equation}
\Lambda_0^{^1S_0-^3P_0}+0.67\Lambda_1^{^1S_0-^3P_0}+0.43\Lambda_0^{^3S_1-^1P_1}+
0.29\Lambda_1^{^3S_1-^3P_1}=
661\pm 169~.
\end{equation}
Note that this constraint is quite similar to that obtained from $\vec{\mathrm{p}}+^4$He: both systems
involve an unpaired proton.  As with $^{18}$F, the nuclear mixing matrix element used in the
analysis was determined from the axial-charge $\beta$ decay of $^{19}$Ne, linking the
same states (up to an isospin rotation).   The details of this determination can be found
in \cite{Adelberger:1983zz}.
\end{enumerate}

The plot omits other constraints either because their interpretation is uncertain, or because
the measurement that have been done lack the precision needed to place meaningful
constraints on hadronic PNC.  Examples of the former include the anapole moment measurement
for $^{133}$Cs \cite{Wood:1997zq} and the circular polarization of the 2.789 MeV $\gamma$ ray emitted in the
decay of $^{21}$Ne \cite{Snover:1978zz,Earle:1983ji}.  The interpretations of these experiments depend on shell model
estimates of quite complicated polarizabilities (Cs) or suppressed mixing matrix elements ($^{21}$Ne),
and there are reasons to believe the associated errors could be large \cite{Haxton:2013aca}.
The plot also omits hadronic PNC constraints established
in experiments on neutron spin rotation in $^4$He \cite{Snow:2011zza},
on the longitudinal analyzing power for the capture of polarized protons on deuterium \cite{Nagle:1978vn},
on the circular polarization
of $\gamma$-rays produced in the capture of unpolarized neutrons on deuterium $\mathrm{nd} \rightarrow \mathrm{t} \gamma$ \cite{Kny84},
and on the $\gamma$-ray asymmetry from the capture of polarized neutrons on deuterium \cite{Avenier:1984is} .
The first three experiments established upper bounds that are not sufficiently restrictive to
impact our analysis.  The fourth experiment produced a signal much larger than expected, and
for that reason is widely thought to reflect an unidentified experimental systematic.

Yet even with the restriction to four experiments -- $\vec{\mathrm{p}}+\mathrm{p}$, $\vec{\mathrm{p}}+^4$He, $P_\gamma(^{18}\mathrm{F})$, and $A_\gamma(^{19}\mathrm{F})$ -- it is clear that all five partial-wave amplitudes contribute to
at least one of the observables.  Early versions of Fig. \ref{fig:standard} arose because it was
noted that the observables plotted depended on very similar combinations of DDH isoscalar and
isovector couplings -- the combination $h_\rho^0 +0.7 h_\omega^0$ and effectively $h_\pi^1$, with
very small corrections due to $h_\rho^1$ and $h_\omega^1$.   While isotensor PNC contributes
to $A_L(\vec{\mathrm{p}} \mathrm{p})$, it plays no role in the other observables.  As the purpose of the plot is to
determine whether there is consistency among competing experiments, it thus made sense to
``freeze out" this degree of freedom.   Thus was done by ``marginalizing" over the isotensor
contribution, allowing it to vary over the DDH reasonable range, while fitting the isoscalar
coupling to the measure $A_L(\vec{\mathrm{p}} \mathrm{p})$.  This procedure effectively expands the allowed $\mathrm{pp}$
band in Fig. \ref{fig:standard} -- though this expansion is modest because DDH assigned
a relatively small reasonable-range uncertainty on $h_\rho^2$ of $ \pm 20$\%.

We want to make several observations about Fig \ref{fig:standard}:
\begin{enumerate}
\item  The figure shows a region of good overlap between the four experiments.  Over most of the last 15 years
similar plots showed some tension among the experiments, but it was found recently \cite{Haxton:2013aca} that 
this largely originated from the use of inconsistent sets
of strong interaction couplings in past analyses.  As PNC experimental observables
are bilinear in the weak and strong couplings, a 
fixed set of strong couplings must be used when extracting values of the DDH
weak couplings.  When the inconsistencies were corrected in \cite{Haxton:2013aca},
the $\mathrm{pp}$ error band move upward, increasing the region of overlap with the results from $\vec{\mathrm{p}}+^4$He
and $^{19}$F.  This is a welcome development, indicating that patterns are beginning to emerge
from experiments.
\item  Another motivation for Fig. \ref{fig:standard} was to emphasize that certain combinations
of weak couplings dominate most observables: thus in some leading-order sense, identifying patterns
in hadronic PNC may not require a five-dimensional analyses.   EFT approaches provide no guidance
on such issues.  In contrast, it is clear that certain of the DDH
parameters (using their best values as a guide) are more important than others.  Unfortunately,
one of those parameters is $h_\pi^1$ -- undercutting any confidence one might have
in relying on DDH for establishing the hierarchy of S-P couplings.   This raises
the question of whether some other theoretical basis exists for considering certain of the LECs
as leading, and others as less important, allowing us to focus effort initially on establishing the
values of the most important parameters.
\item  Fig. \ref{fig:standard} is 2D -- but the end solution is consistent with $h_\pi^1 \sim 0$.
Only one parameter is needed, the isoscalar strength.  Had one known this at the start, one
might have made another axis choice, using the isoscalar and isotensor directions to define
a plane for displaying the experimental results.
\end{enumerate}

Recently there has emerged from large-$N_c$ analyses a way to classify the S-P LECs
as either leading or sub-leading.  That classification appears to be consistent with the
hadronic PNC phenomenology just described, and provides a sound theoretical argument for focusing
first on a particular 2D cut through the 3D isoscalar-isotensor volume.  In the context of large $N_c$,
the absence of a signal from $^{18}$F can be
seen as important confirmation of an emerging pattern.

We now describe the new classification of LECs that merges from large-$N_c$ QCD, and
how this classification influences our interpretation of existing and anticipated hadronic PNC data.

\section{The Large-$N_c$ Classification and Experimental Implications}
The lovely large-$N_c$ work of \cite{Phillips:2014kna,Schindler:2015nga} motivates us to pivot in the 5D space of LECs $\Lambda$ to
two new principal axes,
one in the 2D $I=0$ plane and one along the $I=2$ direction
\begin{eqnarray}
\Lambda_0^+ \equiv {3 \over 4}\Lambda_{0}^{{}^3S_1-{}^1P_1} +{1 \over 4}\Lambda_{0}^{{}^1S_0-{}^3P_0}  & \sim& N_c\nonumber\\
\Lambda_2^{^1S_0-^3P_0}&\sim& N_c ,\label{eq:mxLO}
\end{eqnarray}
with the remaining three orthogonal axes suppressed in the $1/N_c$ counting
\begin{eqnarray}
\Lambda_0^- \equiv {1 \over 4}\Lambda_{0}^{{}^3S_1-{}^1P_1} -{3 \over 4}\Lambda_{0}^{{}^1S_0-{}^3P_0} &\sim& 1/N_c \nonumber \\
\Lambda_1^{^1S_0-^3P_0}&\sim&\sin^2\theta_w\nonumber\\
\Lambda_1^{^3S_1-^3P_1}&\sim&\sin^2\theta_w . \label{eq:mxNNLO}
\end{eqnarray}
One can consider these three subdominant directions to be N$^2$LO, contributing only in relative order $ \sim 1/N_c^2$:
this is explicitly the case for the second isoscalar axis, while for the
$\Delta I=1$ amplitudes additional suppression is gained from the weak mixing angle,
$\sin^2\theta_w/N_c \, \sim\, 12$.  Consequently the large-$N_c$ classification may prove to be especially useful
in hadronic PNC, with corrections only at the $\sim$ 10\% level.

\begin{table}[h]
\caption{A large-$N_c$ hadronic PNC ``Rosetta stone":  The LECs for the S-P PNC potential of Eq. (\ref{eq:kj})
are organized according to the large-$N_c$ classification of \cite{Schindler:2015nga}.  The relationships to the
DDH potential and to the coefficients of Girlanda's EFT potential are shown.
Note that multiplicative factor of  $2m_N m_\rho^2$  must be applied to the
Girlanda entries to obtain the dimensionless coefficients $\Lambda$, e.g.,
$\Lambda_{1}^{{}^1S_0-{}^3P_0}={\cal G}_2~[2 m_N m_\rho^2] $.}
\label{tab:ros}
\begin{center}
\begin{tabular}{cccc}
\hline
& & & \\[-.3cm]
Coeff & DDH & Girlanda & Large $N_c$  \\[.2cm]
\hline
 & & & \\[-.2cm]
$\Lambda_0^+\equiv{3 \over 4}\Lambda_{0}^{{}^3S_1-{}^1P_1} +{1 \over 4}\Lambda_{0}^{{}^1S_0-{}^3P_0}$ &$ - g_\rho h_\rho^0 ({1 \over 2}$+${5 \over 2} \chi_\rho) -g_\omega h_\omega^0({1 \over 2}$-${1 \over 2}\chi_\omega) $ & $2{\cal G}_1 +\tilde{{\cal G}}_1$ & $\sim N_c $\\ [.2cm]
$\Lambda_0^-\equiv{1 \over 4}\Lambda_{0}^{{}^3S_1-{}^1P_1} -{3 \over 4}\Lambda_{0}^{{}^1S_0-{}^3P_0}$&  $g_\omega h_\omega^0 ({3 \over 2}+\chi_\omega)+{3 \over 2}g_\rho h_\rho^0 $ &$ -{\cal G}_1-2\tilde{{\cal G}}_1 $& $\sim {1/N_c}$ \\[.2cm]
$\Lambda_{1}^{{}^1S_0-{}^3P_0}$ & $-g_\rho h_\rho^1 (2$+$ \chi_\rho)- g_\omega h_\omega^1 (2$+$\chi_\omega)$ & ${\cal G}_2 $& $ \sim \sin^2{\theta_w}$\\[.2cm]
$\Lambda_{1}^{{}^3S_1-{}^3P_1} $&  $  {1 \over \sqrt{2}}  g_{\pi NN} h_\pi^1 \left( {m_\rho \over m_\pi} \right)^2+ g_\rho (h_\rho^1-h_\rho^{1\prime})-g_\omega h_\omega^1$  & $2{\cal G}_6 $& $ \sim \sin^2{\theta_w}$  \\[.2cm]
$\Lambda_{2}^{{}^1S_0-{}^3P_0} $ & $ -g_\rho h_\rho^2 (2+ \chi_\rho)$ & $-2 \sqrt{6}{\cal G}_5 $& $\sim N_c$ \\[.2cm]
\hline
\end{tabular}
\end{center}
\end{table}

Table \ref{tab:ros} is a large-$N_c$ version of the ``Rosetta stone" table of \cite{Haxton:2013aca}.  For the Girlanda
coefficients, the key relationships are ${\cal G}_1 \sim \tilde{{\cal G}}_1 \sim N_c$ and ${\cal G}_1 +2 \tilde{{\cal G}}_1 \sim 1/N_c$ \cite{Schindler:2015nga}.   The relationships to the DDH parameters are also shown.  On computing DDH best-value equivalents and comparing them to large-$N_c$ expectations, one finds
\begin{equation}
\left\{ \begin{array}{c} {}^\mathrm{DDH}\Lambda_0^+ \\[0.2cm] {}^\mathrm{DDH}\Lambda_{2}^{{}^1S_0-{}^3P_0} \end{array} \right\}  = \left\{\begin{array}{c} 319 \\[0.2cm] 151 \end{array} \right\} ~~~~~~~~\left\{ \begin{array}{c} {}^\mathrm{DDH}\Lambda_0^- \\[0.1cm] {}^\mathrm{DDH}\Lambda_{1}^{{}^1S_0-{}^3P_0} \\[0.1cm]{}^\mathrm{DDH}\Lambda_{1}^{{}^3S_1-{}^3P_1}  \end{array} \right\} = \left\{ \begin{array}{c} -70 \\[0.1cm] 21 \\[0.1cm] 1340 \end{array} \right\},
\label{eq:DDHcomp}
\end{equation}
with the LO contributions on the left and the corrections on the right. The units are $10^{-7}$.
There is a glaring discrepancy in the $\Lambda_{1}^{{}^3S_1-{}^3P_1}$ isovector channel, where the pion contributes.
The DDH value for $\Lambda_0^-$ is also not negligible.

\subsection{Experimental constraints on large-$N_c$ LECs}
In addition to the above results, we expect to have a new constraint from NPDGamma in hand soon.
NPDGamma data taking is finished and the statistical uncertainty of the 
result has been given as approximately 13 ppb \cite{Fry16}.  Current efforts are focused on measuring and subtracting potential systematic
effects, including an asymmetry associated with 
aluminum in the target window.  Consequently we express
the anticipated asymmetry as
\begin{equation}
\, |A_\gamma| < \epsilon \,1.3 \times 10^{-8}
\end{equation}
under the conservative assumption that the result will be an upper bound (it need not be so) which we set at the statistical
uncertainty, while including a parameter $\epsilon>1$ that will account for consequences of systematic errors, including that associated with
the aluminum subtraction.
We then find \cite{Desplanques:1980pa,Haxton:2013aca} (see also \cite{Liu:2006dm,deVries:2015pza})
\begin{equation}
|\Lambda_1^{^3S_1-^3P_1}|<\epsilon \, 270 ~. \label{eq:npdg}
\end{equation}
The numerical coefficient provides a measure of the potential impact of the result, given the anticipated statistical error.
This bound is important because it is approximately as restrictive as that from $P_\gamma(^{18}$F), but
has a different dependence on the LECs. 

\begin{figure}[t]
\centerline{\includegraphics[width=\textwidth]{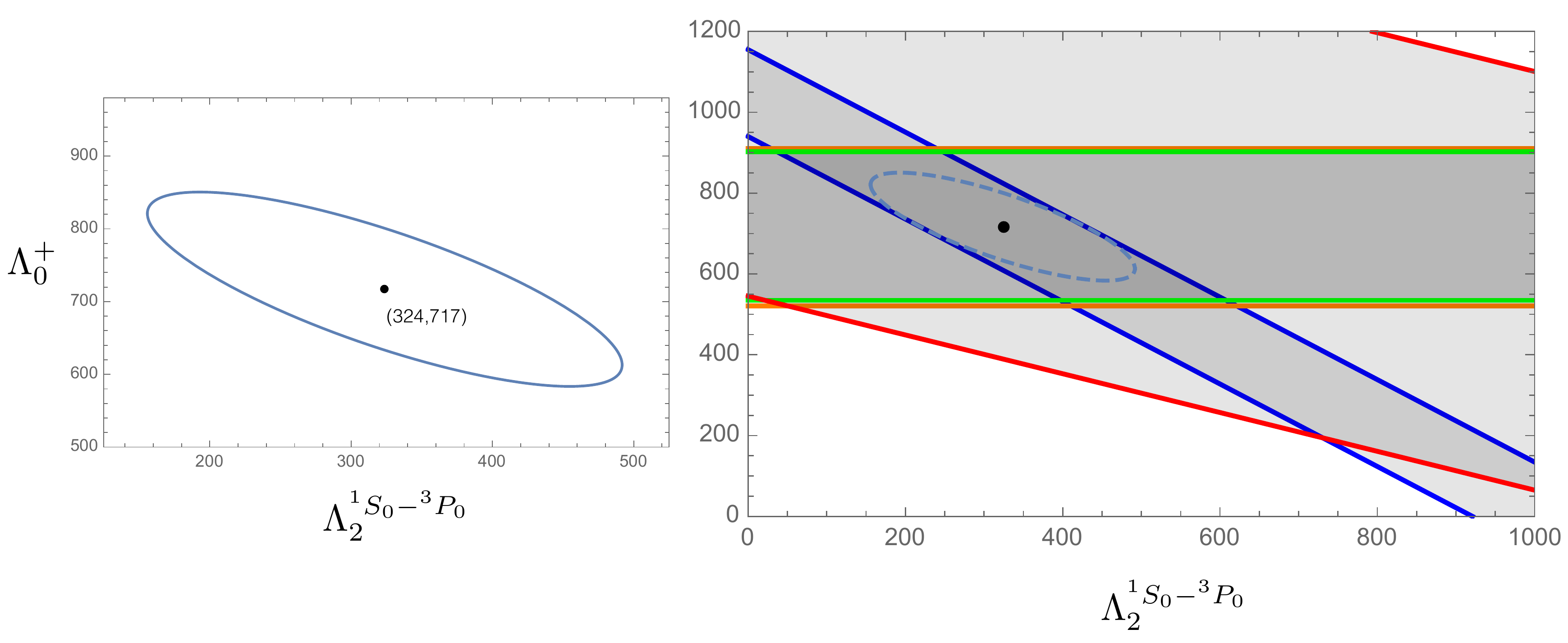}}
\caption{LO large-$N_c$ solutions satisfying all low-energy constraints on hadronic PNC.  The left
panel provides an expanded view of the region, interior to the ellipse, with $\chi^2<1$.
The dot marks the best-fit point.  On the right the constraints from $A_L(\vec{\mathrm{p}}\mathrm{p})$
at low energies (blue boundary),  $A_L(\vec{\mathrm{p}}\mathrm{p})$ at 221 MeV (red),
$A_L(\vec{\mathrm{p}} \alpha)$ (orange), and $A_\gamma(^{19}\mathrm{F})$ (green) are
shown, along fit the combined allowed region (dashed ellipse).  The experimental bands are $1\sigma$. The LECs are given in
units of $10^{-7}$.}
\label{fig:LOFit}
\end{figure}

We now express all five results discussed above in the large-$N_c$ LEC basis,
sequestering the N$^2$LO terms in brackets
\begin{eqnarray}
{2 \over 5} \Lambda_0^+ +  {1 \over \sqrt{6}} \Lambda_2^{{}^1S_0-{}^3P_0} + \left[ -{6 \over 5} \Lambda_0^- +\Lambda_1^{{}^1S_0-{}^3P_0}\right] &=&419 \pm 43 \hspace{0.8cm} A_L(\vec{\mathrm{p}}\mathrm{p}) \nonumber \\
1.3 \Lambda_0^+ + \left[ -0.9 \Lambda_0^-+0.89\Lambda_1^{{}^1S_0-{}^3P_0}+0.32 \Lambda_1^{{}^3S_1-{}^3P_1} \right]&=&930 \pm 253 \hspace{0.65cm} A_L(\vec{\mathrm{p}} \alpha)\nonumber \\
\left[ |2.42 \Lambda_1^{{}^1S_0-{}^3P_0} + \Lambda_1^{{}^3S_1-{}^3P_1}| \right]&<& 340 \hspace{1.7cm} P_\gamma(^{18}\mathrm{F}) \nonumber \\
0.92 \Lambda_0^+ +\left[-1.03 \Lambda_0^- +0.67 \Lambda_1^{{}^1S_0-{}^3P_0} +0.29 \Lambda_1^{{}^3S_1-{}^3P_1} \right]&=& 661 \pm 169 \hspace{0.6cm} A_\gamma(^{19}\mathrm{F}) \nonumber \\
\left[ |\Lambda_1^{{}^3S_1-{}^3P_1}| \right]&<& \epsilon \, 270 \hspace{1.4cm} A_\gamma(\vec{\mathrm{n}}\mathrm{p} \rightarrow \mathrm{d} \gamma)~.
\label{eq:NNLO}
\end{eqnarray}
The LO approximation corresponds to ignoring the bracketed terms while solving
the three remaining equations for $\Lambda_0^+$ and $\Lambda_2^{{}^1S_0-{}^3P_0}$.
The best-value solution is $\Lambda_0^+= 717$ and $\Lambda_2^{{}^1S_0-{}^3P_0}=324$, with a nearly vanishing $\chi^2$ (reflecting
the almost exact overlap of the $A_L(\vec{\mathrm{p}} \alpha)$ and $A_\gamma(^{19}\mathrm{F})$ bands).  The
contour of $\chi^2=1$ (the fit has one degree of freedom) encloses the region
shown in Fig. \ref{fig:LOFit}.  

These best values are both more than a factor of two larger than the DDH benchmark 
values for $\Lambda_0^+$ and $\Lambda_2^{{}^1S_0-{}^3P_0}$ given in
Eq. (\ref{eq:DDHcomp}).  This indicates that there may be a second shortcoming in Fig. \ref{fig:standard}, from the 
perspective of large-$N_c$ QCD:  not only were the wrong isospin axes used, but the marginalization that was
done to remove the effects of $\Lambda_2^{{}^1S_0-{}^3P_0}$ from the band for  $A_L(\vec{\mathrm{p}} \mathrm{p})$
likely underestimated the
associated uncertainties.  In the procedures leading to Fig. \ref{fig:standard} it was assumed that 
the value for $h_\rho^2$, and consequently  $\Lambda_2^{{}^1S_0-{}^3P_0}$, 
would be good to within the estimate reasonable range
of $\pm$ 20\% around the best value.  But the best-value value we found is far outside this band.  In fact most
of the allowed region for $\Lambda_2^{{}^1S_0-{}^3P_0}$ within the ellipse of Fig. \ref{fig:LOFit} would have
also been excluded from this band.
Consequently it is not surprising that there is a
discrepancy between the isoscalar parameter employed in Fig. \ref{fig:standard}, $-(h_\rho^0 +0.7 h_\omega^0)$,
and that associated with $\Lambda_0^+$, $-(h_\rho^0+0.2 h_\omega^0)$.

  It is also apparent that there is no evidence for any
nonzero contribution from the three N$^2$LO LECs.  One way to illustrate this
is to solve the three equalities above, turning on just one of the three N$^2$LO
LECs, while using the central experimental values.  This
yields in turn for the three choices
\begin{eqnarray}
(\Lambda_0^+,\Lambda_2^{{}^1S_0-{}^3P_0}, \Lambda_0^-) &=& (710,309,-7) \nonumber \\
(\Lambda_0^+,\Lambda_2^{{}^1S_0-{}^3P_0}, \Lambda_1^{{}^1S_0-{}^3P_0}) &=& (667,199,71) \nonumber \\
(\Lambda_0^+,\Lambda_2^{{}^1S_0-{}^3P_0}, \Lambda_1^{{}^3S_1-{}^3P_1}) &=& (704,336,45) .\nonumber
\end{eqnarray}
In two cases the result LO parameters do not move outside the $\chi^2=1$ ellipse
of Fig. \ref{fig:LOFit}, showing that any sensitivity to N$^2$LO parameters is
buried under experimental noise.   As the values of the N$^2$LO parameters are
typically comparable
to or less than the experimental uncertainties, we conclude that with current data
\[ \Lambda_0^- \sim \Lambda_1^{{}^1S_0-{}^3P_0} \sim \Lambda_1^{{}^3S_1-{}^3P_1} \sim 0 .\]
Thus we do not have the number or quality of results needed to place meaningful
constraints on parameters we expect to be $\sim 1/10$ the strength of the LO
parameters.  This helps to put the NPDGamma effort in context: it represents a heroic
effort to make the first measurement of a N$^2$LO LEC.

\subsection{The TRIUMF 221 MeV $\boldsymbol{A_L}(\vec{\boldsymbol{\mathrm{p}}}$p)}
We noted earlier that the DDH potential is equivalent to an EFT near threshold, and
thus can be viewed as a model for extrapolating those results to higher momenta, where
ranges controlled by meson masses become important, and where additional degrees
of freedom connected with P-D and other high partial wave amplitudes play a role.  Conversely,
we can work this process in reverse.   The two axes of the ellipse in the left panel of Fig. \ref{fig:standard}
correspond to two constraints, one corresponding to the S-P component of the scattering and the
second from the P-D component.  The former we have already treated in the analysis of the
low-energy data on $A_L(\vec{\mathrm{p}}\mathrm{p})$, but the latter is additional.  From
the second of Eqs. (32) of \cite{Haxton:2013aca} we obtain the constraint
\begin{equation}
 \Lambda_0^+ +  0.48 \Lambda_2^{{}^1S_0-{}^3P_0} + \left[ 2.03 \Lambda_0^- +18.8(h_\rho^1-h_\omega^1) \right] =1063 \pm 518 \hspace{0.8cm} A_L(\vec{\mathrm{p}}\mathrm{p})\big|_{221 \mathrm{~MeV}}~.
 \end{equation}
 Note that a new isovector term arises, expressed in terms of DDH couplings, reflecting the fact
 that P-D scattering includes new degrees of freedom.
For our just determined best values for $\Lambda_0^+$ and $\Lambda_2^{{}^1S_0-{}^3P_0}$,
the LO contribution
to the left-hand side is 873.  There is no evidence for any nonzero N$^2$LO contribution.

\subsection{Deconstructing $h_\pi^1$: A ``$\Delta I=0$'' Rule from Theory and Experiment}
We have already noted that there is tension between the values we have found for $\Lambda_0^+$
and $\Lambda_2^{{}^1S_0-{}^3P_0}$ and DDH best values, with the large-$N_c$ LO LECs
being about a factor of two large than their DDH analogs.  While 
uncertainties in our fit allow for values of $\Lambda_2^{{}^1S_0-{}^3P_0}$
as small as the DDH best value, such small values come at the cost of increasing $\Lambda_0^+$ even further,
as Fig. \ref{fig:LOFit} shows.  Thus such an adjustment does not remove the tension.

A even larger discrepancy exists for $\Lambda_{1}^{{}^3S_1-{}^3P_1}$,
as one can see by comparing the large-$N_c$ predictions of Eqs.~(\ref{eq:mxLO},\ref{eq:mxNNLO})
with that of the DDH best values of Eq.~(\ref{eq:DDHcomp}). The
DDH best value is at least an order of magnitude
larger than the naive large-$N_c$ expectation for an N$^2$LO parameter.  This is
the LEC to which one-pion exchange and thus $h_\pi^1$ contributes.  
The purpose of this section is to describe the likely origin of this mismatch.

Equation~(\ref{eq:mn}) shows that the numerical value of
${}^\mathrm{DDH}\Lambda_{1}^{{}^3S_1-{}^3P_1}$ is dominated by the pion, as the
contribution from vector meson terms is less than 1\% of the total.
This comes in part because the pion-propagator
at low momentum transfers generates
a relative enhancement of $(m_\rho^2/m_\pi^2) \sim 30$.
This magnifies the pion contribution to the ${}^\mathrm{DDH}\Lambda_{1}^{{}^3S_1-{}^3P_1}$ LEC,
and also enhances this LEC relative to others.
If we compare ${}^\mathrm{DDH} \Lambda^{{}^3S_1-{}^3P_1}$ to ${}^\mathrm{DDH} \Lambda_0^+$,
no other large, distinguishing factors are found:
for example,  the DDH best-value
effective pion coupling, $g_{\pi \mathrm{NN}} h_\pi^1/\sqrt{32} \sim 1.08$, is comparable
to those appearing in isoscalar channels,
$-g_\rho h_\rho^0/2 \sim 1.59$ and $-g_\omega h_\omega^0/2 \sim 0.80$.
Thus one is led to the conclusion that a small value for ${}^\mathrm{DDH}\Lambda_{1}^{{}^3S_1-{}^3P_1}$
consistent with the large-$N_c$ hierarchy, requires a significant reduction in the DDH best value for
$h_\pi^1$ -- by a factor of $10$ or more -- to
compensate for the propagator enhancement.

The anatomy of the $h_\pi^1$ coupling involves
both the charged and neutral weak currents. An estimate for the charged current,
or ``Cabibbo'' term, can be based on SU(3)$_f$ symmetry, either
with~\cite{mckellar1967one,fischbach1968application,tadic1968weak,Kummer:1968ra}
or without~\cite{Fischbach:1973vg,Box:1974wa} the use of PCAC and current algebra techniques.
This contribution to the matrix element for $\mathrm{n}\rightarrow \mathrm{p} \pi^-$ can
be related to known hyperon decays~\cite{fischbach1968application}
\begin{equation}
n_-^0 = - \left( \frac{2}{3}\right)^{1/2} \left[ 2 \Lambda_-^0 - \Xi_-^- \right]
\tan \theta_c \equiv g_\pi ~.
\label{nsumrule}
\end{equation}
Recalling that $\tan\theta_c = V_\mathrm{us} /V_\mathrm{ud}$ and using the CKM fit of
Eq.~(12.27) in the CKM review of Ref.~\cite{patrignani2016review}
we find $\tan \theta_c = 0.231$.
Using the experimental values in Table 6.3 of Ref.~\cite{donoghue1986low},
but the phase
conventions of Appendix B in Ref.~\cite{Desplanques:1979hn},
we find $g_\pi = 0.376 \times 10^{-7}$.

This charged-current contribution can be compared to the
``best value'' of $h_\pi^1$ from Ref.~\cite{Desplanques:1979hn},
$h_\pi^1 \sim 12 g_\pi$, the value that was used in Eq. (\ref{eq:DDHcomp}).
The difference reflects in part the neutral current contribution, but also
numerical factors associated with the renormalization group (RNG) evolution of
the operators of the effective Hamiltonian to the
low-momentum scale of interest.  There are also SU(3)$_f$ breaking effects to consider.
Before discussing these effects explicitly we note that the
methods that yield the so-called sum rule estimate of Eq.~(\ref{nsumrule})
can also be used to compute $n_0^0$, the amplitude for $n \rightarrow n \pi^0$. In this case, were CP not broken, we should find
$n_0^0=0$, as first noted by Barton~\cite{barton1961notes}. Indeed this cancellation
has been demonstrated explicitly~\cite{Desplanques:1979hn}, and this
serves as a consistency check of the
method.

Turning to the balance of the terms in the $\Delta S=0$ effective weak Hamiltonian, we
see, unfortunately, that the methods used in the charged current sector cannot
be applied because they rely on the
the existence of  $(V-A)\times(V-A)$ structures in
the weak Hamiltonian. In this case
the needed $\pi \mathrm{NN}$ matrix elements have been
estimated using an factorization ansatz when possible, supplemented by
quark model estimates of the non-factorizable contributions.
This, along with an estimate of the leading RNG evolution effects,
yields a best estimate of $h_\pi^1 = (0.5 + 11.5)g_\pi = 12 g_\pi$, where the separated
factors indicate the charged and neutral current contribution, respectively~\cite{Desplanques:1979hn}.

However, this analysis has been
revisited, particularly by Dubovik and Zenkin [DZ]~\cite{Dubovik:1986pj},
who extended the three-flavor analysis of DDH to include charm quarks and thus the
possibility of GIM cancellation in loop effects~\cite{glashow1970weak}.
They also evaluate non-factorizable
contributions within the MIT bag model. Including operator mixing
and RNG evolution,  they  find a smaller value of $h_\pi^1 \sim 4g_\pi$, but
comparable values of $h_\rho^1$ and $h_\omega^1$~\cite{Dubovik:1986pj}. The possibility of
$\Delta$ resonance contributions has been explored by Feldman et al.~\cite{Feldman:1991tj} in the
DDH framework, and they have revisited $h_\pi^1$ as well, modifying the computation
of the factorized contribution as well as that of
the strong enhancement associated with the charged-current
contribution to find $h_\pi^1 \sim 7 g_\pi$.
The updates made by \cite{Dubovik:1986pj} and \cite{Feldman:1991tj}
cannot be easily combined.
We regard $h_\pi^1 \sim 1.3 \times 10^{-7}$~\cite{Dubovik:1986pj} as the better estimate.

The computations of $h_\pi^1$ we have considered thus far
work within a constituent quark model framework,
so that operators with strange quarks, albeit with no net strangeness, play no role.
We consider model approaches that can address such operators
as well.   The possibility of a considerable enhancement
within the context of factorization and dimensional analysis has been noted~\cite{kaplan1993analysis}.
This issue has been addressed in the Skyrme model. The two-flavor Skyrme model with
vector mesons yields a small value of the order of
$h_\pi^1 \sim 0.3\times 10^{-7}$~\cite{Kai892}, and thus comparable
to $g_\pi$. A three-flavor Skyrme model, which can incorporate
empirical baryon masses, magnetic moments, and hyperon decays
fairly well, has been used to assess the role of four-fermion operators with
$({\bar q} q)({\bar s} s)$ flavor structure, yielding values
for $h_\pi^1$ in the range $(0.8- 1.3)\times 10^{-7}$, considerably larger
than the two-flavor result~\cite{Meissner:1998pu}. The authors of
Ref. \cite{Meissner:1998pu}
stress that this result is not a consequence of a large (scalar) strangeness component
in the nucleon wave function, a notion now in disfavor
due to lattice QCD results~\cite{Freeman:2012ry,Oksuzian:2012rzb,Junnarkar:2013ac,Gong:2013vja}, but rather
of operators appearing that involve strange quarks.  It is worth noting, however, that the effective PNC Hamiltonian
they employ at a scale of  $\sim$ 1 GeV includes neither QCD renormalization nor mixing effects in
evolving from the weak scale. Such effects would presumably be muted, as noted by DZ \cite{Dubovik:1986pj},
by the GIM effects that can arise when the charm quark is included and could lead to additional
cancellations.  We note that a computation of the low-energy effective Hamiltonian including LO QCD evolution 
with heavy quarks has been made in the $\theta_c=0$ limit~\cite{Dai:1991bx}; this is
not a good approximation, however, in the $\Delta I=1$ sector because the neutral
current contribution enters with a factor of $\sin^2 \theta_w /3 \simeq 0.08$. 
For reference the $|\Delta S|=1$ low-energy effective Hamiltonian, with heavy quarks, 
has been computed to NLO in QCD~\cite{Buchalla:1995vs}.  



To conclude this section we note that improved assessments of $h_\pi^1$ yield
results that are considerably smaller than the ``best estimate'' given by DDH, and
are less incompatible with the picture suggested by large $N_c$ and naive
dimensional analysis.

The emerging picture points to a dominance of the $\Delta I =0$ hadronic PNC
nucleon amplitudes relative to $\Delta I=1$ ones, resulting from a combination
of the suppression of the latter, and a not insignificant enhancement of the
former, relative to DDH best values.
This is reminiscent of the dominance of
the $\Delta I=1/2$ amplitude in $K\to \pi\pi$ decay,
as was originally pointed out in \cite{Adelberger:1985ik}.  We note the recent
work of \cite{boyle2013emerging,Buras:2014maa}.

\section{Next Steps}
The above arguments show that five significant experimental constraints on hadronic PNC -- $A_L(\vec{\mathrm{p}} \mathrm{p})$ near
threshold and at 221 MeV, $A_L(\vec{\mathrm{p}}\alpha)$, $P_\gamma(^{1}$F), and $A_\gamma(^{19}$F) --
are in excellent accord with expectations arising
from large-$N_c$ QCD.   In particular, this approach provides two leading-order parameters $\Lambda_0^+$ and $\Lambda_2^{{}^1S_0-{}^3P_0}$
that can be determined from the existing experiments, and appear to account very well for all observations.
Three S-P N$^2$LO LECs, expected to be $\sim 10\%$ of the LO LECs, do appear to be small: at least, no existing experiment
requires assigning a nonzero value to any of the non-leading LECs.  Large $N_c$ thus provides a hierarchical simplification
of standard EFT approaches, breaking the five degrees of freedom into a 2+3 pattern.  As we have discussed, such a simplification
has been tried before, but without sound motivation, leading to descriptions that in hindsight appear flawed.  In contrast,
the large-$N_c$ hierarchy appears to be sound from the vantage points of both theory and experiment, and particularly useful for
hadronic PNC because the correction terms are N$^2$LO, not NLO.  The approach  provides us with a sensible starting point for
planning future work to refine our characterization of the hadronic weak interaction, and thus to understand how this interaction is modified
when embedded in strongly interacting systems.

\subsection{Testing the LO Theory}
Despite the quality of the LO fit, there is not a lot of redundancy, especially with the constraints from
$A_L(\vec{\mathrm{p}} \alpha)$ and $A_\gamma(^{19}$F) being so similar.   Thus an additional independent measurement
sensitive to the LO couplings would be valuable.  Furthermore, while the value of $A_L(\vec{\mathrm{p}} \mathrm{p})$ is known
to 10\%, the errors on the other two experiments exceed 25\%.   A new measurement matching
the precision of $A_L(\vec{\mathrm{p}} \mathrm{p})$, but probing a different combination of $\Lambda_0^+$ and
$\Lambda_2^{{}^1S_0-{}^3P_0}$, thus could substantially shrink the allowed ellipse shown in Fig. \ref{fig:LOFit}.
A more precise determination of the LO LECs
would be important for future searches for N$^2$LO LECs: in experiments where these terms arise in combination with LO terms,
even modest errors in LO parameters would obscure the effects of N$^2$LO corrections.
There do appear to be opportunities to generate new, high quality constraints on the LO parameters.\\

\begin{figure}[t]
\centerline{\includegraphics[width=\textwidth]{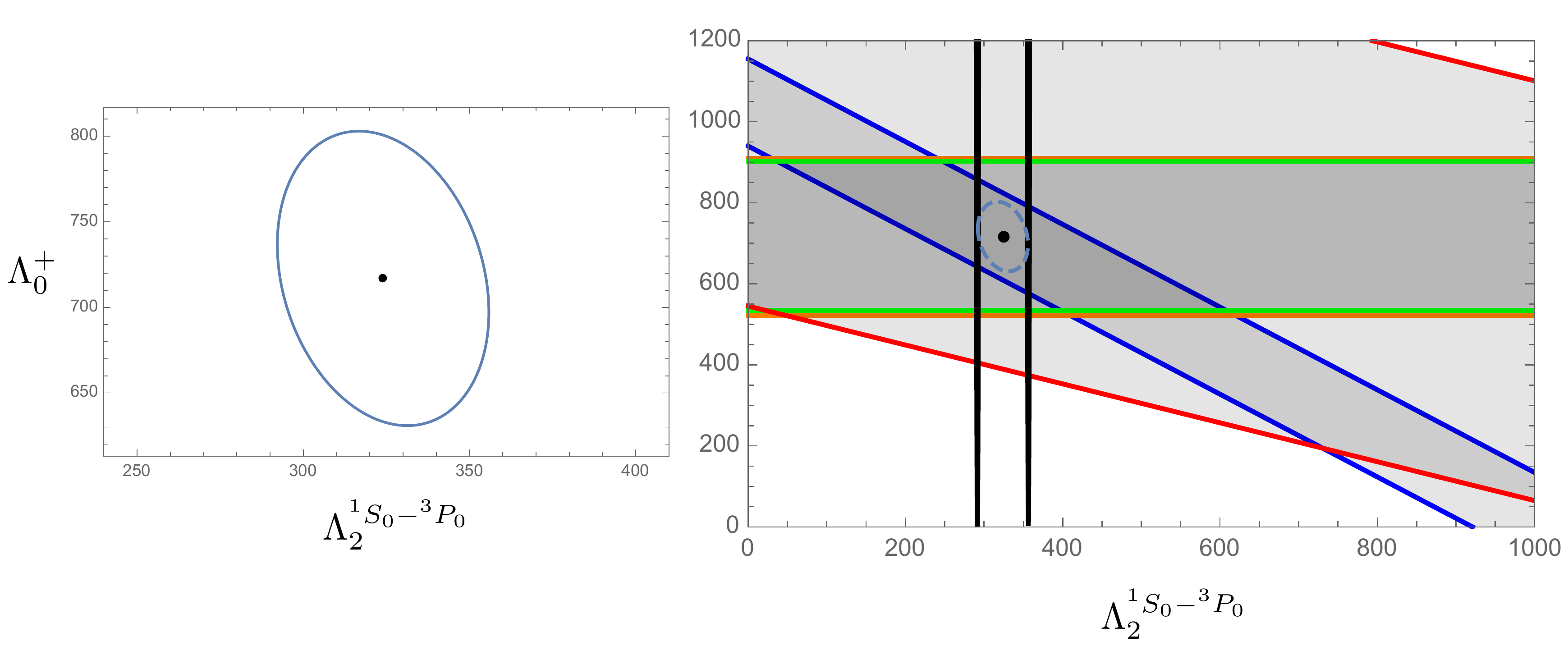}}
\caption{As in Fig. \ref{fig:LOFit}, but adding the impact of a future LQCD calculation of the $\Delta I=2$ amplitude $\Lambda_2^{{}^1S_0-{}^3P_0}$ to $\pm$ 10\%, centered on the central value from Fig. \ref{fig:LOFit}.}
\label{fig:LOFitLQCD}
\end{figure}

\noindent
{\it Lattice QCD:}
In lattice QCD (LQCD) one solves strongly interacting problems by replacing the continuum problem with a 
discretized version, a finite grid in Euclidean space-time with periodic boundary conditions.  While this precludes
any direct calculation of scattering amplitudes \cite{Sci5}, the distortion of the energy levels in a finite volume can be related
to low-energy scattering parameters \cite{Sci6,Sci7,Sci8} using techniques developed by L\"{u}scher \cite{Sci9,Sci10}.
Most NN scattering calculations documented in the literature were performed with nuclear sources that placed both nucleons
at the same space-time point, limiting the results to s-waves.  In contrast, applications to hadronic PNC, where p-waves are clearly
essential, require the use of extended nuclear sources, placed on the lattice in a variety of configurations that, in sum, allow one
to associate lattice eigenvalues with 
partial waves having good spherical symmetry.  This is a nontrivial problem given the cubic symmetry of the lattice.  
The first calculation of parity-odd two-nucleon
scattering using L\"{u}scher's method were recently performed, demonstrating the technique \cite{Sci36}.

There is an effort underway to apply LQCD to the problem of calculating $\Lambda_2^{{}^1S_0-{}^3P_0}$ \cite{Sci83}.
Because this
scattering amplitude carries $\Delta I=2$, there are no disconnected (quark loop) contributions \cite{Tiburzi:2012xx}.  Thus the statistical noise in this channel should
be significantly lower than in $\Delta I=0,1$ channels, opening up the possibility of a good LQCD ``measurement" near the physical
pion mass.  A calculation of hadronic PNC in the $\Delta I =2$ channel is expected to be an order of magnitude less costly than
a measurement in the $\Delta I=1$ channel.  Preliminary work on $\Delta I=2$ NN PN  yielded a non-zero signal \cite{Sci83}, and led
to the identification of improved interpolating operators, significantly reducing the contamination from nucleon excited states at
early times, at least for the heavy pion masses ($m_\pi \sim 700$ and $\sim 800$ MeV) that were used in these exploratory calculations.
Phenomenologically relevant calculations will likely require calculations with $m_\pi \lesssim 300$ MeV, where valid comparisons can be
made to low-energy EFTs \cite{Sci85,Sci86,Sci87,Sci88,Sci89,Sci90}.

The gestation period for major experiments in hadronic PNC, such as NPDGamma, can approach a decade or more.  Recent improvements
in LQCD applications to NN interactions have been rapid, and the basic tools are in place for a major attack on $\Lambda_2^{{}^1S_0-{}^3P_0}$.
Thus LQCD might turn out to be the fastest route to determining $\Lambda_2^{{}^1S_0-{}^3P_0}$ to the desired precision of 10\%.
The impact of such a result, illustrated in Fig. \ref{fig:LOFitLQCD}, would be quite significant.

There was an early, exploratory calculation of $h_\pi^1$ in lattice QCD by Wasem, at $m_\pi \sim 389$ MeV \cite{SciWasem}.
Instead of an NN amplitude, a three-point function \cite{Beane:2002ca} corresponding to a nucleon-to-resonance transition through pion absorption
was calculated.  The calculation did not include nonperturbative renormalization of the bare PNC operators, a chiral extrapolation
to the physical pion mass, or the contributions from disconnected (quark loop) diagrams.  The result obtained,
$h_\pi^1 = (1.10 \pm 0.51 \pm 0.06) \time 10^{-7}$, is consistent with the $^{18}$F upper bound $|h_\pi^1| \lesssim 1.3 \times 10^{-7}$.
The recent developments describe above, for direct calculations of NN PNC amplitudes, now supersede this approach.\\

\noindent
{\it New experiments constraining LO LECs:}  While one could envision developing new experiments to complement existing
measurements on $A_L(\vec{\mathrm{p}}\mathrm{p})$, it strikes us that the most conservative strategy might be to re-examine
previous efforts on $A_L(\vec{\mathrm{p}} \alpha)$ and $A_\gamma(^{19}$F), to see if improvements are possible.
Specifically, our analysis uses results from the 1985 measurement of
$A_L(\vec{\mathrm{p}} \alpha)=(-3.34 \pm 0.9) \times 10^{-7}$ from Lang et al. \cite{Lang:1985jv}, which was
performed with a 1.3 $\mu$A polarized beam from the Swiss Institute for Nuclear Research cyclotron.  The beam's polarization was switched
at the ion source every 30 msec.  The experiment utilized techniques that the group had developed in its earlier $A_L(\vec{\mathrm{p}}\mathrm{p})$
experiment \cite{Balzer:1980dn,Balzer:1985au,Kistryn:1987tq}, to control systematics.  If the 1$\sigma$ error bar on this result could be reduced by a factor $\sim 2.5$, the desired
precision of $\sim$ 10\% would be reached.

From the error budget provided, and assuming that the statistical and systematic errors were added in quadrature in forming the
final result, it appears that the statistical contribution is somewhat larger.  Thus an experiment delivering about an order
of magnitude more beam on target, combined with a reduction in the systematic error of about a factor of two, could be required.
The latter will be challenging: the PSI group worked very hard to measure and correct
for residual transverse components in the polarized beam, induced by nonuniform magnetic fields in the cyclotron.  This was
the principal systematic.  Thus
a new experiment would need to do even better.  Nevertheless, it strikes us that an approach such as this that builds on past
experience, with a proven technology and with
sources of systematic error well documented -- could be preferable to starting an effort that lacks such a history.

The few-body theory used to relate the measurements to NN S-P amplitudes should also be updated: there are several modern
techniques that could be applied to this problem, including quantum Monte Carlo \cite{Nollett}.

Alternatively, one could consider a new attempt on $A_\gamma(^{19}$F), one of two cases where axial charge $\beta$ decay is
available as a nuclear matrix element calibration.  One important aspect of this experiment is the success
in controlling systematic errors:  The analysis of \cite{Elsener:1987sx} found that systematic errors were negligible, contributing
to the overall uncertainty at a level that would have allowed a 10\% measurement, had the statistics been available.  Thus a repetition
of this experiment with a factor of 20 increase in counting could reach the target 10\% uncertainty.  As $^{19}$F was produced
using a relatively modest 0.4 $\mu$A 5 MeV polarized proton beam, the needed statistics might be attainable.

\subsection{Testing the N$^2$LO Theory: NPDGamma and $\boldsymbol{P_\gamma(^{18}}$F)}
A significant outcome of our work is the recognition that 1) past experiments have done a good job in characterizing
the LO large-$N_c$ interaction -- with further improvements possible in the near term, such as that illustrated
in Fig. \ref{fig:LOFitLQCD} -- and 2) we have already embarked on a credible campaign to learn
about the N$^2$LO corrections.  From the perspective of this second point, the striking aspect of Eqs. (\ref{eq:NNLO})
is that $\Lambda_1^{{}^1S_0-{}^3P_0}$ and and $\Lambda_1^{{}^3S_1-{}^3P_1}$ are the low-hanging fruit in this
endeavor, because we can use isospin to restrict ourselves to the $\Delta I=1$ plane in our 5D parameter space, where
no LO terms exist to mask the smaller effects we seek.  Furthermore, we have already embarked on a nearly optimal
program to limit or measure these parameters, with $P_\gamma(^{18}$F) and NPDGamma being ideal choices for this task.

\begin{figure}[t]
\centerline{\includegraphics[width=0.7\textwidth]{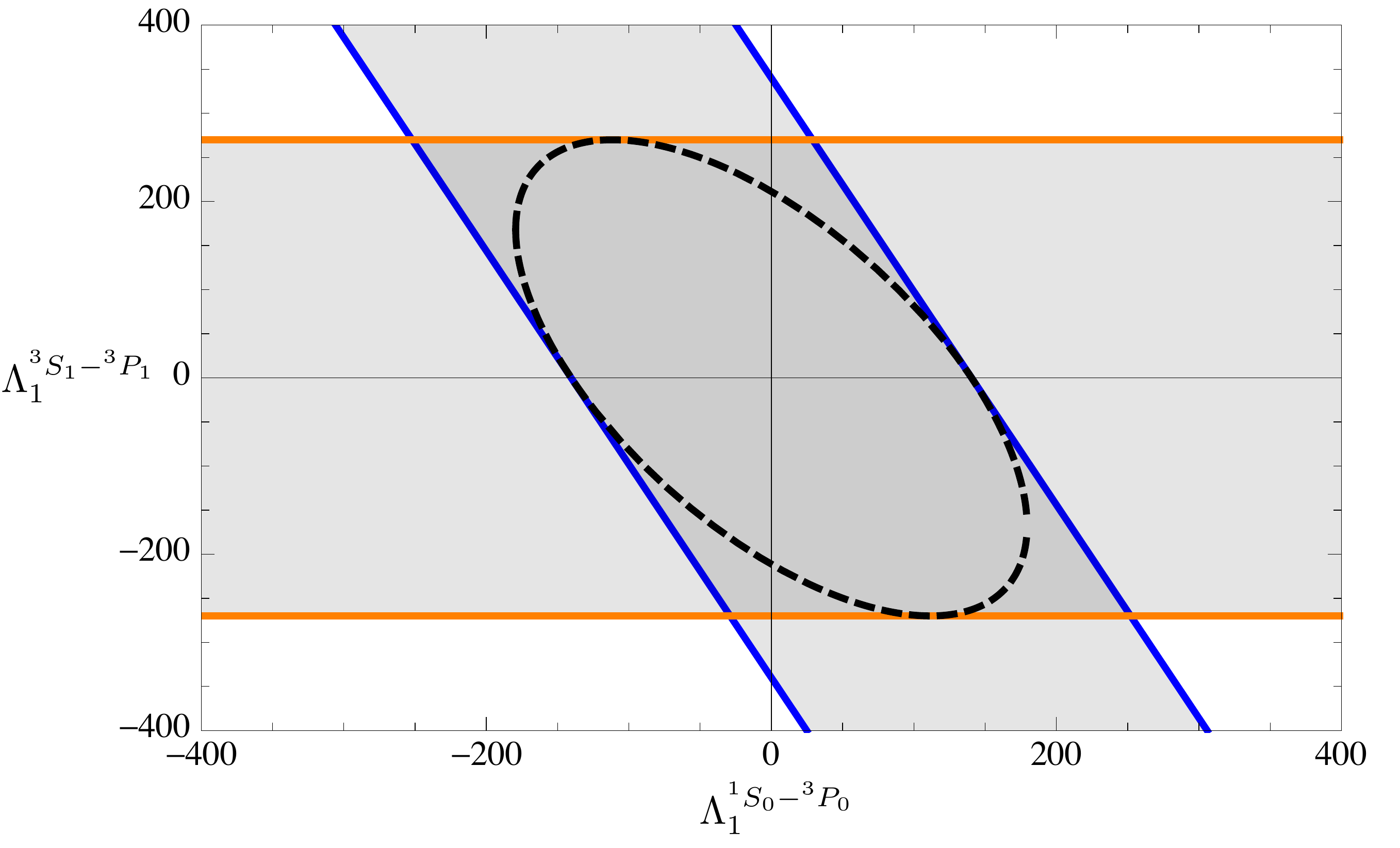}}
\caption{The progress in constraining the large-$N_c$ isovector N$^2$LO LECs that will result from
combining anticipated NPDGamma results (horizontal band) with the existing
constraint from $P_\gamma(^{18}$F) (vertical band).  The former assumes a central value of zero
for $A_\gamma(\vec{\mathrm{n}}\mathrm{p} \rightarrow \mathrm{d} \gamma)$ and an uncertainty determined by the experiment's
statistics, and thus assumes that the current campaign to subtract out window-induced asymmetries will
yield a final systematic uncertainty well below the statistical uncertainty.  Note that both isovector LECs are bounded once
NPDGamma results are combined with $P_\gamma(^{18}$F), while neither is bounded without this result.
The LECs are in
units of $10^{-7}$.}
\label{fig:NNLO}
\end{figure}

An important question to ask is where we might stand, once NPDGamma announces its result.  To assess this we make the
choice $\epsilon \sim 1$, which is a possible outcome as the important systematic effects in the experiment appear to
be isolated in the window subtraction, including Al as the dominant correction.  The net results that would follow
from combining the bound on $P_\gamma(^{18}$F) with
a NPDGamma $A_\gamma(\vec{\mathrm{n}}\mathrm{p} \rightarrow \mathrm{d} \gamma)$ result centered on zero with a final error bar of $1.3 \times 10^{-8}$  is shown in Fig. \ref{fig:NNLO}.  Note that a central value for
$A_\gamma$ other than zero would shift the horizontal
band up or down, while significant residual systematic uncertainties leading to $\epsilon>1$ would broaden the band proportionately.
One observes that the two experiments are very complementary, probing different combinations of the two $\Delta I =1$
LECs.  If one uses $\Lambda_0^+ \sim 700$ as the scale of the LO contribution, then current $P_\gamma(^{18})$F and potential NPDGamma
constraints are about a factor of three below the LO scale, or roughly at the NLO level.  Thus considerable work remains ahead, as
these constraints should be improved another factor of three, if the LECs are of natural size.

Figure \ref{fig:NNLO} also shows that the NPDGamma result is essential in placing such bounds on $\Lambda_1^{{}^1S_0-{}^3P_0}$ and
$\Lambda_1^{{}^3S_1-{}^3P_1}$: $P_\gamma(^{18}$F) by itself constrains neither to be below the LO scale.  The two measurements
are highly complementary, testing distinct combinations of the $\Delta I=1$ LECs.

Next-generation experiments to reach the N$^2$LO level will be challenging, and would be more easily motivated if the needed sensitivity
to see N$^2$LO contributions could
be better defined.  If current LQCD efforts to determine $\Lambda_2^{{}^1S_0-{}^3P_0}$ at the 10\% level meet with success,
then the N$^2$LO contributions might be the next LQCD challenge.  While the presence of disconnected diagrams would substantially
increase the difficulty of $\Delta I=1$ calculations, the detection of a signal of any quality would be helpful to the field, potentially
confirming the predicted large-$N_c$ coupling hierarchy and providing a definite target for the experimentalists.

\subsection{The potential impact of new experiments}
Several recent proposals have been made to initiate new experiments, and others have been pursued or discussed in the past.
The large-$N_c$ LEC hierarchy can be used to determine how new experiments will impact our understanding of
hadronic PNC.  We first describe some of the potential observables:
\begin{itemize}
\item[a)] An effort to measure the longitudinal asymmetry for scattering polarized neutrons from a $^3$He target -- $\vec{n}+{}^3{\rm He}\rightarrow {}^3{\rm H}+p$ -- is underway at the SNS.  The predicted value, based on calculations by Viviani et al. using the AV18 potential with the UIX three-body interaction is~\cite{Viviani:2010qt}
    \begin{eqnarray}
    { 364  \over 10^{-8}}\,A_\mathrm{p}=-\Lambda_0^++0.227 \Lambda_2^{{}^1S_0-{}^3P_0}- \left[ 3.82 \Lambda_0^-
    +8.18\Lambda_1^{{}^1S_0-{}^3P_0}+2.27\Lambda_1^{{}^3S_1-{}^3P_1} \right]~.
    \end{eqnarray}
 Using the best-fit parameters for the two LO LECs yields
 $A_\mathrm{p} \sim -1.8 \times 10^{-8}$.  As the relative sign of the two LO LECs is opposite to that found for $A_L(\vec{\mathrm{p}} \mathrm{p})$,
 a measurement of a nonzero value would generate a complementary band in Fig. \ref{fig:LOFit}, testing the region of intersection
 identified there.

\item[b)]  There was a past attempt to detect the gamma ray asymmetry produced when longitudinally polarized neutrons
capture on a deuterium target, $\vec{\mathrm{n}}+\mathrm{d}\rightarrow \mathrm{t}+\gamma$.  $A_\gamma(\vec{\mathrm{n}}\mathrm{d})$ is a rare example of a few-nucleon PNC observable where
some enhancement occurs, as the parity-conserving M1 amplitude is suppressed at thermal neutron energies.  The predicted asymmetry is~\cite{Desplanques:1986cq}
    \begin{eqnarray}
    { 118  \over 10^{-7}}\,A_\gamma=\Lambda_0^++0.44 \Lambda_2^{{}^1S_0-{}^3P_0}- \left[ 1.86 \Lambda_0^-
    +0.65 \Lambda_1^{{}^1S_0-{}^3P_0}+0.42\Lambda_1^{{}^3S_1-{}^3P_1} \right],
    \end{eqnarray}
 yielding $\sim 7.3 \times 10^{-7}$ for the LO best values.   The value obtained in an experiment by Avenier et al.~\cite{Avenier:1984is},
 $\sim 8\times 10^{-6}$, is an order of magnitude larger.  This result has been largely ignored because of concerns about
 unidentified systematics \cite{Adelberger:1985ik}.

\item[c)]  There are several np observables complementary to NPDGamma's $A_\gamma(\vec{\mathrm{n}}\mathrm{p})$.  Significant
effort has been invested in studies of the circular polarization of the 2.2 MeV gamma ray produced in thermal neutron capture on
the proton.  The circular polarization \cite{Desplanques:1980pa},
\begin{equation}
{825  \over 10^{-7}} \, P_\gamma= \Lambda_0^+ +1.27 \Lambda_2^{{}^1S_0-{}^3P_0} + \left[0.47 \Lambda_0^- \right],
\end{equation}
is $\sim 1.4 \times 10^{-7}$ for the LO LEC best values.  The combination of three separate measurements and various
control experiment by Knyaz'kov et al. led to a determination $P_\gamma=(1.8 \pm 1.8) \times 10^{-7}$ \cite{Kny84},
superseding an earlier result that apparently was contaminated by bremmstrahlung from the reactor core \cite{Lobashov:1972fwg}.
Recently interest has been expressed in measuring the inverse reaction, the circular polarization dependence of the
breakup reaction $\vec{\gamma} d \rightarrow \mathrm{n}+\mathrm{p}$, which of course has an identical dependence on the LECs.
This has been proposed as a commissioning experiment at at upgraded Hi$\gamma$s facility~\cite{Snow:2016zyq}

  \item[d)] The neutron spin rotation of polarized cold neutrons traversing a parahydrogen target is a third possible np
  observable.  The experiment is feasible because strong interaction spin-flip scattering off parahydrogen (S=0 molecules) is
  forbidden.  The spin rotation, taken from the Paris potential results of \cite{Avishai:1985mu} as modified by \cite{Heckel1986,Adelberger:1987zt} (see also \cite{Schiavilla:2004wn,
  Liu:2006dm,Griesshammer:2011md}),
  \begin{equation}
  {180 \over 10^{-7}} {d \phi^{\mathrm{n}} \over dz} \Big|_\mathrm{parahydrogen}=\left( \Lambda_0^+ +2.82 \Lambda_2^{{}^1S_0-{}^3P_0} -\left[ 3.15 \Lambda_0^-+1.94 \Lambda_1^{{}^3S_1-{}^3P_1} \right] \right) \mathrm{~rad/m},
  \end{equation}
  is $\sim 9.1 \times 10^{-7}$ rad/m for the best-value LO LECs.  To date there has been no experiment mounted. This is
  also the case with a fourth possible np observable, the dependence of the capture cross section on the
  neutron helicity, which tests a combination of LECs identical to that appearing above.  Note that the neutron spin
  rotation in deuterium has also been calculated \cite{Schiavilla:2008ic}.

\item[e)] A neutron spin rotation experiment has been attempted in He, where from \cite{Dmitriev:1983mg},
\begin{equation}
{105 \over 10^{-7}} {d \phi^{\mathrm{n} } \over dz} \Big|_{{}^4\mathrm{He}}= \left(\Lambda_0^+- \left[ 1.61 \Lambda_0^-
    +0.92 \Lambda_1^{{}^1S_0-{}^3P_0}+0.35\Lambda_1^{{}^3S_1-{}^3P_1} \right] \right) \mathrm{~rad/m}
 \end{equation}
 which leads to a predicted LO spin rotation of $6.8 \times 10^{-7}$ rad/m.  The experiment, performed on the slow neutron
 beam line at NIST, established the upper bound $(1.7 \pm 9.1 \pm 1.4) \times 10^{-7}$ \cite{Snow:2011zza}.  

  \item[f)] The longitudinal analyzing power for the scattering of polarized protons on deuterium, where \cite{Desplanques:1979st}
  \begin{equation}
  {156 \over 10^{-8}} \, A_L = -\Lambda_0^+ + \left[ 1.75 \Lambda_0^- -1.09 \Lambda_1^{{}^1S_0-{}^3P_0} -1.25 \Lambda_1^{{}^3S_1-{}^3P_1} \right],
  \end{equation}
  leading to a LO large-$N_c$ estimate of $A_L = -4.6 \times 10^{-8}$.   An experiment was done at 15 MeV nearly 40 years ago,
  yielding the result $A_L=(-3.5 \pm 8.5) \times 10^{-8}$ \cite{Nagle:1978vn}.

\end{itemize}

\begin{table}[h]
\caption{Candidate future hadronic PNC experiments, including several that have been or are now being pursued.  The LO large-$N_c$
estimates for the observables are given, along with the functional dependence on the LECs. }
\label{tab:newexp}
\begin{center}
\begin{tabular}{|c|c|c|c|}
\hline
& & & \\[-0.4cm]
Observable & Exp. Status & LO Expectation & LO LEC Dependence  \\[.2cm]
\hline
 & & & \\[-.2cm]
 $A_\mathrm{p}$($\vec{\mathrm{n}}+^3\mathrm{He} \rightarrow {}^3$H+p) & ongoing & $-1.8 \times 10^{-8}$ & $-\Lambda_0^++0.227\Lambda_2^{{}^1S_0-{}^3P_0}$ \\[.2cm]
 $A_\gamma(\vec{\mathrm{n}}+\mathrm{d} \rightarrow \mathrm{t}+\gamma$) &$8 \times 10^{-6}{}$ (see text) \cite{Avenier:1984is} & $~~7.3 \times 10^{-7}$  &$ \Lambda_0^++0.44 \Lambda_2^{{}^1S_0-{}^3P_0}$ \\[.2cm]
 $P_\gamma(\mathrm{n}+\mathrm{p} \rightarrow \mathrm{d}+\gamma)$ & $(1.8 \pm 1.8) \times 10^{-7}$ \cite{Kny84} &$ 1.4 \times 10^{-7}$ & $\Lambda_0^+ +1.27 \Lambda_2^{{}^1S_0-{}^3P_0}$ \\[.2cm]
$ {d \phi^{\mathrm{n}} \over dz}\big|_\mathrm{parahydrogen}$ &  none & $ 9.4 \times 10^{-7}$ rad/m & $ \Lambda_0^+ +2.7 \Lambda_2^{{}^1S_0-{}^3P_0}$ \\[.3cm]
${d \phi^{\mathrm{n} } \over dz} \big|_\mathrm{{}^4\mathrm{He}}$ & $(1.7 \pm 9.1 \pm 1.4) \times 10^{-7}$ \cite{Snow:2011zza} & $6.8 \times 10^{-7}$ rad/m &$ \Lambda_0^+ $\\[.3cm]
$A_L(\vec{\mathrm{p}}+\mathrm{d})$ & $(-3.5 \pm 8.5) \times 10^{-8}$ \cite{Nagle:1978vn} & $-4.6 \times 10^{-8}$ & $ -\Lambda_0^+$\\[.2cm]
\hline
\end{tabular}
\end{center}
\end{table}

A summary of this discussion is presented in Table \ref{tab:newexp}.  All of the above experiments belong in the same class as
those displayed in Fig. \ref{fig:LOFit}, presumably dominated by the LO LECs.  Thus it would be appropriate to set goals,
for future efforts on these observables, that will guarantee they improve the pattern in that figure.  To be competitive,
uncertainties should then be achieved that are at least comparable to those of the $A_L(\vec{p} \alpha)$ and $A_\gamma(^{19}$F) measurements,
which as noted previously are $\sim 25\%$.  The last two experiments in the table are only sensitive to $\Lambda_0^+$, and thus can
be viewed as surrogates for $A_L(\vec{p} \alpha)$ and $A_\gamma(^{19}$F).  As we have argued that it is important to
reduce the errors on these measurements to $\sim10\%$, better defining the two LO LECs,
one might consider whether that goal could be reached more easily via $A_L(\vec{\mathrm{p}}\mathrm{d})$ or neutron spin
rotation in $^4$He.

Four of the experiments depend on linear combinations of $\Lambda_0^+$ and $\Lambda_2^{{}^1S_0-{}^3P_0}$,
and in three cases these combinations
are not too different from that tested in $A_L(\vec{\mathrm{p}}\mathrm{p})$.  Consequently not a lot will be
learned unless a $\sim 10\%$ uncertainty comparable to that of $A_L(\vec{\mathrm{p}}\mathrm{p})$ is
achieved.  The observable $A_\mathrm{p}$($\vec{\mathrm{n}}+^3$He $\rightarrow {}^3$H+p), which is the subject of an
ongoing experiment \cite{Crawford16}, is notable because the contributions from $\Lambda_0^+$ and $\Lambda_2^{{}^1S_0-{}^3P_0}$ carry
opposite signs.   A high quality result would place a fifth band on Fig. \ref{fig:LOFit}, oblique to those now there.  That would
very helpful in testing the large-$N_c$ picture.  An uncertainty on
 $A_\mathrm{p}$($\vec{\mathrm{n}}+^3$He $\rightarrow {}^3$H+p) of
$\lesssim 0.5 \times 10^{-8}$ would have a major impact on Fig \ref{fig:LOFit}.

All of these remarks are made under the assumption that N$^2$LO corrections are indeed of the naively expected size, $\sim 10\%$.
If this assumption is wrong, then the question of the role of candidate new experiments becomes more complicated.
We also remark that none of the experiments in Table \ref{tab:newexp} is like $^{18}$F and NPDGamma, exclusively
sensitive to N$^2$LO corrections.  This underscores how important these two experiments are, providing a unique
opportunity to establish the scales of two of the three N$^2$LO LEC cleanly, and thus to verify the predicted large-$N_c$ LEC
hierarchy.

\section{Summary and Outlook}

The field of hadronic PNC began three score years ago, when it was recognized that the weak semileptonic interaction responsible for nuclear beta decay violates parity and that there should exist a corresponding parity-violating signal in the NN interaction due to the nonleptonic weak force. 
By exploiting the parity violation, it was realized one could isolate this interaction despite its embedding in a strongly interacting
system, testing our understanding of W- and Z-exchange among the quarks.  NN and nuclear systems were recognized as uniquely important
to the study of the neutral current, which plays no role in
strangeness-changing interactions due to the absence of flavor-changing neutral currents.
For many years it has been anticipated that the neutral current would be revealed in a PNC $\Delta I=1$ interaction of long range,
generated by pion exchange.  The expectation that this interaction would have important consequences for PNC observables
has influenced PNC analyses since the 1980s.  Yet, the combination of experimental
data, large-$N_c$ arguments, and subsequent re-examinations of QCD effects influencing the size of $h_\pi^1$, today all indicate otherwise.

Experimental progress in this field has been slow due to the difficulty of measuring effects with a natural size of order $(Gm_\pi^2)\sim 10^{-7}$ relative to the strong interaction: despite the distinctive character of PNC observables,
there are many systematics that can lead to false
signals at this level.  On the theoretical side one had the challenge of five independent S-P Danilov amplitudes, and very few measurements
to constrain these amplitudes.  There was a need to find some organizing principle to simplify this task.  Influenced by meson-exchange
models of PNC, the interest in the $\Delta I=1$ interaction, and the absence of $\Delta I=2$ contributions to the data displayed in
Fig. \ref{fig:LOFit} (apart from $A_L(\vec{\mathrm{p}}\mathrm{p})$), the choice was made to focus on average $\Delta I=0$ and 1 strengths, reducing the S-P interaction to two
effective couplings.  Compounded by some inconsistent treatments
of couplings, resolved only recently \cite{Haxton:2013aca}, this led to a puzzling pattern, including the conclusion that the $\Delta I=1$ degree of freedom was unneeded.

Meson exchange models like the DDH potential attempt to predict weak couplings, a daunting task given the uncertainties inherent
in embedding weak interactions in a strongly interacting environment.  Beginning about 10 years ago, pionless effective field theories
began to be employed.  As we have noted in this article, this approach is closely related to the Danilov amplitudes and to the contact
potential of Desplanques and Missimer, which date to the field's early days.  Furthermore the DDH potential is operationally
identical to pionless EFT, in the S-P limit.  Thus it could be argued there is little new in the EFT approach.  On the other hand, EFT
forces one to again confront the issue of five independent S-P amplitudes: there is no organizational principle in EFT for reducing the
number of LECs to a more manageable number.  The LECs appear as equivalent constants that must be determined from experiment.

This is why the recent application of large-$N_c$ to the PNC is such an important step forward: it provides a hierarchical division
of the LECs into two groups, two LO LEC with  $\Delta I=0$ and $\Delta I=2$, and three N$^2$LO LECs that naively are $\sim 10\%$
corrections, two of which carry $\Delta I=1$ and one $\Delta I=0$.   The purpose of this review has been to apply this formalism to the full body of information available on PNC.
The LO LECs -- $\Lambda_0^+$ and $\Lambda_2^{^1S_0-^3P_0}$ -- can be reasonably well determined from existing PNC results,
and the resulting LO large-$N_c$ effective theory accounts for all  of the existing measurements.
We have also argued that the remaining three constants---$\Lambda_1^{{}1S_0-{}^3P_0}$, $\Lambda_1^{{}^3S_1-{}^3P_1}$, and $\Lambda_0^-$ --
do indeed contribute at a suppressed level.  More precisely, if NPDGamma produces a result close to its announced statistical precision,
it can be combined with $P_\gamma(^{18}$F) to constrain $\Lambda_1^{{}1S_0-{}^3P_0}$ and $\Lambda_1^{{}^3S_1-{}^3P_1}$
to values considerably below that of $\Lambda_0^+$.

In retrospect, the field has been very fortunate in its choice of experiments.  $A_L(\vec{\mathrm{p}}\mathrm{p})$, $A_L(\vec{\mathrm{p}}\alpha)$,
and $A_\gamma(^{19}$F) have allowed us to extract the large-$N_c$ LO LECs.  $P_\gamma(^{18}$F) and $A_\gamma(\vec{\mathrm{n}}\mathrm{p} \rightarrow \mathrm{d}\gamma)$ probe the N$^2$LO $\Delta I=1$ plane in complementary ways, potentially allowing us to demonstrate
the LEC hierarchy suggested by large $N_c$ by constraining two of the subdominant LECs.

It strikes us that, after many years during which the theoretical and experimental situations were less clear, we are beginning to
understand the pattern of PNC in NN systems. There are opportunities to make further progress -- to pick experiments that
optimally constrain the LECs and test the patterns predicted by large $N_c$.  These include
\begin{itemize}
\item [i)] A lattice QCD evaluation of the couplings, beginning with measurement of the $\Delta I=2$ parameter $\Lambda_2^{^1S_0-^3P_0}$.
A measurement accurate to 10\% would significant narrow the uncertainties on $\Lambda_0^+$ and $\Lambda_2^{^1S_0-^3P_0}$ .
This calculation is the natural first step for LQCD, as the $\Delta I=2$ amplitude has no contributions from
disconnected (quark loop) diagrams.

\item[ii)] An improved determination of the LO parameters $\Lambda_0^+$ and $\Lambda_2^{^1S_0-^3P_0}$ by a modern and higher precision measurement of the $\vec{p}\alpha$ longitudinal analyzing power and/or the $^{19}$F photon decay asymmetry.

\item[iii)] Alternatively, an improved determination of the LO parameters $\Lambda_0^+$ and $\Lambda_2^{^1S_0-^3P_0}$ by one of the
new experiments listed in Table \ref{tab:newexp}.  We have noted that $A_L(\vec{n}+{}^3{\rm He}\rightarrow {}^3{\rm H}+p)$ would
be a particularly good choice.

\item[iv)] A test of the N$^2$LO parameters $\Lambda_1^{{}^1S_0-{}^3P_0}$, $\Lambda_1^{{}^3S_1-{}^3P_1}$, and $\Lambda_0^-$ by a modern and higher precision $^{18}$F circular polarization measurement and/or a
second-generation $\vec{n}p\rightarrow d\gamma$ experiment that limits statistical and systematic errors to $\lesssim 0.5 \times 10^{-8}$.
\end{itemize}

Undertaking all or some combination of these efforts could confirm the large-$N_c$ picture outlined and provide accurate values
for four of its five LECs.  This in turn could finally demonstrate that the six-decade-long program to understand hadronic PNC
has been successful.

\begin{center}
{\bf\large Acknowledgements}
\end{center}

The authors acknowledge the hospitality of the Kavli Institute for Theoretical Physics and NSF support under award PHY11-25915. The work of
SG was supported by the Department of Energy under award DE-FG02-96ER40989, while that of
WCH was supported by the Department of Energy under awards DE-SC00046548, DE-AC02-98CH10886, and KB0301052 and by the
Simons Foundation.

\newpage

\bibliographystyle{utphys}
\bibliography{PNC_bib}

\providecommand{\href}[2]{#2}\begingroup\raggedright\begin{thebibliography}{100}

\bibitem{Crawford16}
C.~Crawford, ``{A new era in discrete hadronic symmetries},'' 2016.
\newblock {http://online.kitp.ucsb.edu/online/nuclear-c16/ }.

\bibitem{Fry:2016esm}
J.~Fry {\em et~al.}, ``{Status of the NPDGamma experiment},''
\href{http://dx.doi.org/10.1007/s10751-016-1376-4}{{\em Hyperfine Interact.}
  {\bfseries 238} no.~1, (2017) 11}.

\bibitem{Fry16}
J.~Fry {\em et~al.}, ``{Status of the NPDGamma analysis},'' 2015.
\newblock {http://meetings.aps.org/link/BAPS.2015.APR.S6.2 }.

\bibitem{Sci83}
T.~Kurth, E.~Berkowitz, E.~Rinaldi, P.~Vranas, A.~Nicholson, M.~Strother,
  A.~Walker-Loud, and E.~Rinaldi, ``{Nuclear parity violation from lattice
  QCD},'' {\em PoS} {\bfseries LATTICE2015} (2016) 329,
\href{http://arxiv.org/abs/1511.02260}{{\ttfamily arXiv:1511.02260 [hep-lat]}}.

\bibitem{Haxton:2013aca}
W.~Haxton and B.~Holstein, ``{Hadronic parity violation},''
  \href{http://dx.doi.org/10.1016/j.ppnp.2013.03.009}{{\em Prog. Part. Nucl.
  Phys.} {\bfseries 71} (2013) 185--203},
\href{http://arxiv.org/abs/1303.4132}{{\ttfamily arXiv:1303.4132 [nucl-th]}}.

\bibitem{Phillips:2014kna}
D.~R. Phillips, D.~Samart, and C.~Schat, ``{Parity-violating nucleon-nucleon
  force in the 1/$N_c$ Expansion},''
  \href{http://dx.doi.org/10.1103/PhysRevLett.114.062301}{{\em Phys. Rev.
  Lett.} {\bfseries 114} no.~6, (2015) 062301},
\href{http://arxiv.org/abs/1410.1157}{{\ttfamily arXiv:1410.1157 [nucl-th]}}.

\bibitem{Schindler:2015nga}
M.~R. Schindler, R.~P. Springer, and J.~Vanasse, ``{Large-$N_c$ limit reduces
  the number of independent few-body parity-violating low-energy constants in
  pionless effective field theory},''
  \href{http://dx.doi.org/10.1103/PhysRevC.93.025502}{{\em Phys. Rev.}
  {\bfseries C93} no.~2, (2016) 025502},
\href{http://arxiv.org/abs/1510.07598}{{\ttfamily arXiv:1510.07598 [nucl-th]}}.

\bibitem{Donoghue:1992dd}
J.~F. Donoghue, E.~Golowich, and B.~R. Holstein, ``{Dynamics of the standard
  model},'' {\em Camb. Monogr. Part. Phys. Nucl. Phys. Cosmol.} {\bfseries 2}
  (1992) 1--540.
[Camb. Monogr. Part. Phys. Nucl. Phys. Cosmol.35(2014)].

\bibitem{Maiani:2011zz}
L.~Maiani, ``{Universality of the Weak Interactions, Cabibbo theory and where
  they led us},'' \href{http://dx.doi.org/10.1393/ncr/i2011-10072-5}{{\em Riv.
  Nuovo Cim.} {\bfseries 34} (2011) 679--692},
\href{http://arxiv.org/abs/1303.5000}{{\ttfamily arXiv:1303.5000 [hep-ph]}}.

\bibitem{Desplanques:1979hn}
B.~Desplanques, J.~F. Donoghue, and B.~R. Holstein, ``{Unified treatment of the
  parity violating nuclear force},''
\href{http://dx.doi.org/10.1016/0003-4916(80)90217-1}{{\em Annals Phys.}
  {\bfseries 124} (1980) 449}.

\bibitem{Barnes:1978sq}
C.~Barnes, M.~Lowry, J.~Davidson, R.~Marrs, F.~Morinigo, {\em et~al.},
  ``{Search for neutral weak current effects in the nucleus F-18},''
\href{http://dx.doi.org/10.1103/PhysRevLett.40.840}{{\em Phys. Rev. Lett.}
  {\bfseries 40} (1978) 840--843}.

\bibitem{Biz80}
P.~G. Bizetti {\em et~al.}, ``{Search for parity mixing in $^{18}$F},'' {\em
  Lett. Nuovo Cimento} {\bfseries 29} (1980) 167.

\bibitem{Ahrens:1982}
G.~Ahrens, W.~Harfst, J.~Kass, E.~Mason, H.~Schober, G.~Steffens, H.~Waeffler,
  P.~Bock, and K.~Grotz, ``{Search for parity violation in the 1081 keV gamma
  transition of ${}^{18}$F},''
  \href{http://dx.doi.org/10.1016/0375-9474(82)90280-9}{{\em Nucl. Phys.}
  {\bfseries A390} (1982) 486--508}.

\bibitem{Bini:1985zz}
M.~Bini, T.~Fazzini, G.~Poggi, and N.~Taccetti, ``{Search for the circular
  polarization of the 1081-keV gamma ray in F-18},''
\href{http://dx.doi.org/10.1103/PhysRevLett.55.795}{{\em Phys. Rev. Lett.}
  {\bfseries 55} (1985) 795--798}.

\bibitem{Page:1987ak}
S.~Page, H.~Evans, G.~Ewan, S.~Kwan, J.~Leslie, {\em et~al.}, ``{Weak
  pion-nucleon coupling strength: New constraint from parity mixing in
  $^{18}$F},''
\href{http://dx.doi.org/10.1103/PhysRevC.35.1119}{{\em Phys. Rev.} {\bfseries
  C35} (1987) 1119--1131}.

\bibitem{Haxton:1981sf}
W.~Haxton, ``{Parity nonconservation in $^{18}$F and meson exchange
  contributions to the axial charge operator},''
\href{http://dx.doi.org/10.1103/PhysRevLett.46.698}{{\em Phys. Rev. Lett.}
  {\bfseries 46} (1981) 698}.

\bibitem{Bennett80}
{C. Bennett, M. M. Lowrey, and K. Krien,} {\em Bull. Am. Phys. Soc.} {\bfseries
  25} (1980) 486.

\bibitem{Adelberger:1983zz}
E.~Adelberger, M.~Hindi, C.~Hoyle, H.~Swanson, R.~Von~Lintig, {\em et~al.},
  ``{Beta decays of Ne-18 and Ne-19 and their relation to parity mixing in F-18
  and F-19},''
\href{http://dx.doi.org/10.1103/PhysRevC.27.2833}{{\em Phys. Rev.} {\bfseries
  C27} (1983) 2833--2856}.

\bibitem{Adelberger:1985ik}
E.~G. Adelberger and W.~C. Haxton, ``{Parity violation in the nucleon-nucleon
  Interaction},''
\href{http://dx.doi.org/10.1146/annurev.ns.35.120185.002441}{{\em Ann. Rev.
  Nucl. Part. Sci.} {\bfseries 35} (1985) 501--558}.

\bibitem{Gericke:2011zz}
M.~Gericke, R.~Alarcon, S.~Balascuta, L.~Barron-Palos, C.~Blessinger, {\em
  et~al.}, ``{Measurement of parity-violating gamma-ray asymmetry in the
  capture of polarized cold neutrons on protons},''
\href{http://dx.doi.org/10.1103/PhysRevC.83.015505}{{\em Phys. Rev.} {\bfseries
  C83} (2011) 015505}.

\bibitem{Page05}
S.~A. Page {\em et~al.}, ``{Measurement of parity violation in np capture: the
  NPDGamma experiment},'' {\em J. Res. Natl. Inst. Stand. Technol.} {\bfseries
  110} (2005) 195--204.

\bibitem{Crawford}
C.~Crawford, ``{The NPDGamma experiment at the SNS FnPB},'' 2007.
\newblock {http://meetings.aps.org/link/BAPS.2007.DNP.CD.5 }.

\bibitem{Zhu:2004vw}
S.-L. Zhu, C.~Maekawa, B.~Holstein, M.~Ramsey-Musolf, and U.~van Kolck,
  ``{Nuclear parity-violation in effective field theory},''
  \href{http://dx.doi.org/10.1016/j.nuclphysa.2004.10.032}{{\em Nucl. Phys.}
  {\bfseries A748} (2005) 435--498},
\href{http://arxiv.org/abs/nucl-th/0407087}{{\ttfamily arXiv:nucl-th/0407087
  [nucl-th]}}.

\bibitem{Girlanda:2008ts}
L.~Girlanda, ``{On a redundancy in the parity-violating 2-nucleon contact
  Lagrangian},'' \href{http://dx.doi.org/10.1103/PhysRevC.77.067001}{{\em Phys.
  Rev.} {\bfseries C77} (2008) 067001},
\href{http://arxiv.org/abs/0804.0772}{{\ttfamily arXiv:0804.0772 [nucl-th]}}.

\bibitem{Phillips:2008hn}
D.~R. Phillips, M.~R. Schindler, and R.~P. Springer, ``{An
  effective-field-theory analysis of low-energy parity-violation in
  nucleon-nucleon scattering},''
  \href{http://dx.doi.org/10.1016/j.nuclphysa.2009.02.011}{{\em Nucl. Phys.}
  {\bfseries A822} (2009) 1--19},
\href{http://arxiv.org/abs/0812.2073}{{\ttfamily arXiv:0812.2073 [nucl-th]}}.

\bibitem{Danilov:1965}
G.~Danilov, ``{Circular polarization of $\gamma$ quanta in absorption of
  neutrons by protons and isotopic structure of weak interactions},'' {\em
  Phys. Lett.} {\bfseries 18} (1965) 40--41.

\bibitem{Danilov:1971fh}
G.~Danilov, ``{Dispersion approach to the investigation of the weak
  nucleon-nucleon interaction},''
\href{http://dx.doi.org/10.1016/0370-2693(71)90292-9}{{\em Phys. Lett.}
  {\bfseries B35} (1971) 579--580}.

\bibitem{Danilov:1972}
{G. S. Danilov,} {\em Sov. J. Nucl. Phys.} {\bfseries 14} (1972) 443.

\bibitem{Lee:1956qn}
T.~D. Lee and C.-N. Yang, ``{Question of parity conservation in weak
  Interactions},''
\href{http://dx.doi.org/10.1103/PhysRev.104.254}{{\em Phys. Rev.} {\bfseries
  104} (1956) 254--258}.

\bibitem{Wu:1957my}
C.~S. Wu, E.~Ambler, R.~W. Hayward, D.~D. Hoppes, and R.~P. Hudson,
  ``{Experimental test of parity conservation in beta decay},''
\href{http://dx.doi.org/10.1103/PhysRev.105.1413}{{\em Phys. Rev.} {\bfseries
  105} (1957) 1413--1414}.

\bibitem{Tanner:1957zz}
N.~Tanner, ``{Parity in nuclear reactions},''
\href{http://dx.doi.org/10.1103/PhysRev.107.1203}{{\em Phys. Rev.} {\bfseries
  107} (1957) 1203--1204}.

\bibitem{Haeberli:1995uz}
W.~Haeberli and B.~R. Holstein, ``{Parity violation and the nucleon-nucleon
  system},''
\href{http://arxiv.org/abs/nucl-th/9510062}{{\ttfamily arXiv:nucl-th/9510062
  [nucl-th]}}.

\bibitem{RamseyMusolf:2006dz}
M.~J. Ramsey-Musolf and S.~A. Page, ``{Hadronic parity violation: A new view
  through the looking glass},''
  \href{http://dx.doi.org/10.1146/annurev.nucl.54.070103.181255}{{\em Ann. Rev.
  Nucl. Part. Sci.} {\bfseries 56} (2006) 1--52},
\href{http://arxiv.org/abs/hep-ph/0601127}{{\ttfamily arXiv:hep-ph/0601127
  [hep-ph]}}.

\bibitem{Krane:1971zz}
K.~Krane, C.~Olsen, J.~R. Sites, and W.~Steyert, ``{Observation of 1.5\%
  parity-nonconserving gamma-ray asymmetry},''
\href{http://dx.doi.org/10.1103/PhysRevLett.26.1579}{{\em Phys. Rev. Lett.}
  {\bfseries 26} (1971) 1579--1581}.

\bibitem{Yuan:1991zz}
V.~Yuan, C.~Bowman, J.~Bowman, J.~E. Bush, P.~Delheij, {\em et~al.}, ``{Parity
  nonconservation in polarized-neutron transmission through La-139},''
\href{http://dx.doi.org/10.1103/PhysRevC.44.2187}{{\em Phys. Rev.} {\bfseries
  C44} (1991) 2187--2194}.

\bibitem{Dubovik:1986pj}
V.~M. Dubovik and S.~V. Zenkin, ``{Formation of parity nonconserving nuclear
  forces in the standard model SU(2)(l) X U(1) X SU(3)(c)},''
\href{http://dx.doi.org/10.1016/0003-4916(86)90021-7}{{\em Annals Phys.}
  {\bfseries 172} (1986) 100--135}.

\bibitem{Feldman:1991tj}
G.~B. Feldman, G.~A. Crawford, J.~Dubach, and B.~R. Holstein, ``{Delta
  contributions to the parity violating nuclear interaction},''
\href{http://dx.doi.org/10.1103/PhysRevC.43.863}{{\em Phys. Rev.} {\bfseries
  C43} (1991) 863--874}.

\bibitem{Schindler:2013yua}
M.~Schindler and R.~Springer, ``{The theory of parity violation in few-nucleon
  systems},'' \href{http://dx.doi.org/10.1016/j.ppnp.2013.05.002}{{\em Prog.
  Part. Nucl. Phys.} {\bfseries 72} (2013) 1--43},
\href{http://arxiv.org/abs/1305.4190}{{\ttfamily arXiv:1305.4190 [nucl-th]}}.

\bibitem{Desplanques:1976mt}
B.~Desplanques and J.~H. Missimer, ``{An analysis of parity violating nuclear
  effects at low energy},''
\href{http://dx.doi.org/10.1016/0375-9474(78)96134-1}{{\em Nucl. Phys.}
  {\bfseries A300} (1978) 286--312}.

\bibitem{SciWasem}
J.~Wasem, ``{Lattice QCD calculation of nuclear parity violation},''
  \href{http://dx.doi.org/10.1103/PhysRevC.85.022501}{{\em Phys. Rev.}
  {\bfseries C85} (2012) 022501},
\href{http://arxiv.org/abs/1108.1151}{{\ttfamily arXiv:1108.1151 [hep-lat]}}.

\bibitem{Carlson:2001ma}
J.~Carlson, R.~Schiavilla, V.~Brown, and B.~Gibson, ``{Parity violating
  interaction effects 1: The longitudinal asymmetry in pp elastic
  scattering},'' \href{http://dx.doi.org/10.1103/PhysRevC.65.035502}{{\em Phys.
  Rev.} {\bfseries C65} (2002) 035502},
\href{http://arxiv.org/abs/nucl-th/0109084}{{\ttfamily arXiv:nucl-th/0109084
  [nucl-th]}}.

\bibitem{Eversheim:1991tg}
P.~Eversheim, W.~Schmitt, S.~Kuhn, F.~Hinterberger, P.~von Rossen, {\em
  et~al.}, ``{Parity violation in proton proton scattering at 13.6-MeV},''
\href{http://dx.doi.org/10.1016/0370-2693(91)90209-9}{{\em Phys. Lett.}
  {\bfseries B256} (1991) 11--14}.

\bibitem{Nagle:1978vn}
D.~Nagle, J.~Bowman, C.~Hoffman, J.~McKibben, R.~Mischke, {\em et~al.},
  ``{Parity violation in the scattering of 15-{MeV} protons by hydrogen},''
{\em AIP Conf.Proc.} {\bfseries 51} (1979) 224--230.

\bibitem{Balzer:1980dn}
R.~Balzer, R.~Henneck, C.~Jacquemart, J.~Lang, M.~Simonius, {\em et~al.},
  ``{Measurement of parity nonconservation in pp scattering at 45-MeV},''
\href{http://dx.doi.org/10.1103/PhysRevLett.44.699}{{\em Phys. Rev. Lett.}
  {\bfseries 44} (1980) 699--702}.

\bibitem{Balzer:1985au}
R.~Balzer, R.~Henneck, C.~Jacquemart, J.~Lang, F.~Nessi-Tedaldi, {\em et~al.},
  ``{Parity violation in proton-proton scattering at 45-MeV},''
\href{http://dx.doi.org/10.1103/PhysRevC.30.1409}{{\em Phys. Rev.} {\bfseries
  C30} (1984) 1409--1430}.

\bibitem{Kistryn:1987tq}
S.~Kistryn, J.~Lang, J.~Liechti, T.~Maier, R.~Muller, {\em et~al.},
  ``{Precision measurement of parity violation in proton proton scattering at
  45-{MeV}},''
\href{http://dx.doi.org/10.1103/PhysRevLett.58.1616}{{\em Phys. Rev. Lett.}
  {\bfseries 58} (1987) 1616}.

\bibitem{Berdoz:2002sn}
A.~R. Berdoz {\em et~al.}, ``{Parity violation in proton proton scattering at
  221-MeV},'' \href{http://dx.doi.org/10.1103/PhysRevC.68.034004}{{\em Phys.
  Rev.} {\bfseries C68} (2003) 034004},
\href{http://arxiv.org/abs/nucl-ex/0211020}{{\ttfamily arXiv:nucl-ex/0211020
  [nucl-ex]}}.

\bibitem{Berdoz:2001nu}
A.~R. Berdoz {\em et~al.}, ``{Parity violation in proton proton scattering at
  221 MeV},'' \href{http://dx.doi.org/10.1103/PhysRevLett.87.272301}{{\em Phys.
  Rev. Lett.} {\bfseries 87} (2001) 272301},
\href{http://arxiv.org/abs/nucl-ex/0107014}{{\ttfamily arXiv:nucl-ex/0107014
  [nucl-ex]}}.

\bibitem{Lang:1985jv}
J.~Lang, T.~Maier, R.~Muller, F.~Nessi-Tedaldi, T.~Roser, {\em et~al.},
  ``{Parity nonconservationin elastic p-alpha scattering and the determination
  of the weak meson-nucleon coupling constants},''
\href{http://dx.doi.org/10.1103/PhysRevLett.54.170}{{\em Phys. Rev. Lett.}
  {\bfseries 54} (1985) 170--173}.

\bibitem{Henneck:1982cc}
R.~Henneck, C.~Jacquemart, J.~Lang, R.~Muller, T.~Roser, {\em et~al.}, ``{Study
  of parity nonconservation in p-alpha scattering},''
\href{http://dx.doi.org/10.1103/PhysRevLett.48.725}{{\em Phys. Rev. Lett.}
  {\bfseries 48} (1982) 725--728}.

\bibitem{Elsener:1987sx}
K.~Elsener, W.~Grubler, V.~Konig, P.~A. Schmelzbach, J.~Ulbricht, B.~Vuaridel,
  D.~Singy, C.~Forstner, and W.~Z. Zhang, ``{Parity nonconservation in $^{19}$F
  nuclei},''
\href{http://dx.doi.org/10.1016/0375-9474(87)90411-8}{{\em Nucl. Phys.}
  {\bfseries A461} (1987) 579--602}.

\bibitem{Elsener:1984vp}
K.~Elsener, W.~Grubler, V.~Konig, P.~A. Schmelzbach, J.~Ulbricht, D.~Singy,
  C.~Forstner, W.~Z. Zhang, and B.~Vuaridel, ``{Constraints on weak
  meson-nucleon coupling from parity nonconservation in F-19},''
\href{http://dx.doi.org/10.1103/PhysRevLett.52.1476}{{\em Phys. Rev. Lett.}
  {\bfseries 52} (1984) 1476--1479}.

\bibitem{Wood:1997zq}
C.~Wood, S.~Bennett, D.~Cho, B.~Masterson, J.~Roberts, {\em et~al.},
  ``{Measurement of parity nonconservation and an anapole moment in cesium},''
\href{http://dx.doi.org/10.1126/science.275.5307.1759}{{\em Science} {\bfseries
  275} (1997) 1759--1763}.

\bibitem{Snover:1978zz}
K.~A. Snover, R.~von Lintig, E.~G. Adelberger, H.~E. Swanson, T.~A. Trainor,
  A.~B. McDonald, E.~D. Earle, and C.~A. Barnes, ``{Upper limit on parity
  mixing in Ne-21},''
\href{http://dx.doi.org/10.1103/PhysRevLett.41.145}{{\em Phys. Rev. Lett.}
  {\bfseries 41} (1978) 145--148}.

\bibitem{Earle:1983ji}
E.~D. Earle, A.~B. Mcdonald, E.~G. Adelberger, K.~A. Snover, H.~E. Swanson,
  R.~Von~Lintig, H.~B. Mak, and C.~A. Barnes, ``{Parity mixing in Ne-21:
  Evidence for weak neutral currents in nuclei},''
\href{http://dx.doi.org/10.1016/0375-9474(83)90021-0}{{\em Nucl. Phys.}
  {\bfseries A396} (1983) 221C--230C}.

\bibitem{Snow:2011zza}
W.~Snow, C.~Bass, T.~Bass, B.~Crawford, K.~Gan, {\em et~al.}, ``{Upper bound on
  parity-violating neutron spin rotation in He-4},''
\href{http://dx.doi.org/10.1103/PhysRevC.83.022501}{{\em Phys. Rev.} {\bfseries
  C83} (2011) 022501}.

\bibitem{Kny84}
V.~A. Knyaz'kov, E.~A. Kolomenskii, V.~A. Lobashov, V.~A. Nazarenko, A.~N.
  Pirozhkov, {\em et~al.}, ``{A new experimental study of the circular
  polarization of np capture gamma rays},'' {\em Nucl. Phys.} {\bfseries A417}
  (1984) 209--230.

\bibitem{Avenier:1984is}
M.~Avenier, J.~F. Cavaignac, D.~H. Koang, B.~Vignon, R.~Hart, and R.~Wilson,
  ``{Parity violation in n-d capture},''
\href{http://dx.doi.org/10.1016/0370-2693(84)91119-5}{{\em Phys. Lett.}
  {\bfseries B137} (1984) 125--128}.

\bibitem{Desplanques:1980pa}
B.~Desplanques, ``{Parity nonconserving nuclear forces},''
\href{http://dx.doi.org/10.1016/0375-9474(80)90174-8}{{\em Nucl. Phys.}
  {\bfseries A335} (1980) 147--167}.

\bibitem{Liu:2006dm}
C.~P. Liu, ``{Parity-Violating Observables of Two-Nucleon Systems in Effective
  Field Theory},'' \href{http://dx.doi.org/10.1103/PhysRevC.75.065501}{{\em
  Phys. Rev.} {\bfseries C75} (2007) 065501},
\href{http://arxiv.org/abs/nucl-th/0609078}{{\ttfamily arXiv:nucl-th/0609078
  [nucl-th]}}.

\bibitem{deVries:2015pza}
J.~de~Vries, N.~Li, U.-G. Mei{\ss}ner, A.~Nogga, E.~Epelbaum, and N.~Kaiser,
  ``{Parity violation in neutron capture on the proton: Determining the weak
  pion-nucleon coupling},''
  \href{http://dx.doi.org/10.1016/j.physletb.2015.05.074}{{\em Phys. Lett.}
  {\bfseries B747} (2015) 299--304},
\href{http://arxiv.org/abs/1501.01832}{{\ttfamily arXiv:1501.01832 [nucl-th]}}.

\bibitem{mckellar1967one}
B.~McKellar, ``{The one pion exchange contribution to the weak parity violating
  nucleon-nucleon potential},'' {\em Phys. Lett. B} {\bfseries 26} (1967)
  107--108.

\bibitem{fischbach1968application}
E.~Fischbach, ``{Application of current algebra and partially conserved
  axial-vector current to the weak B B $\pi$ vertex},'' {\em Phys. Rev.}
  {\bfseries 170} (1968) 1398.

\bibitem{tadic1968weak}
D.~Tadi{\'c}, ``{Weak parity-nonconserving potentials},'' {\em Phys. Rev.}
  {\bfseries 174} (1968) 1694.

\bibitem{Kummer:1968ra}
W.~Kummer and M.~Schweda, ``{An SU(6) estimate of the effective nonleptonic
  parity-violating coupling in strangeness-conserving weak processes},''
{\em Acta Phys. Austriaca} {\bfseries 28} (1968) 303--308.

\bibitem{Fischbach:1973vg}
E.~Fischbach and D.~Tadic, ``{Parity violating nuclear interactions and models
  of the weak Hamiltonian},''
  \href{http://dx.doi.org/10.1016/0370-1573(73)90025-2}{{\em Phys. Rept.}
  {\bfseries 6} (1973) 123--186}.

\bibitem{Box:1974wa}
M.~A. Box, B.~H.~J. McKellar, P.~Pick, and K.~R. Lassey, ``{Parity mixing in
  nuclei as a test of theories of the weak interaction},''
  \href{http://dx.doi.org/10.1088/0305-4616/1/5/003}{{\em J. Phys.} {\bfseries
  G1} (1975) 493}.

\bibitem{patrignani2016review}
C.~Patrignani, P.~D. Group, {\em et~al.}, ``{Review of particle physics},''
  {\em Chinese Phys. C} {\bfseries 40} (2016) 100001.

\bibitem{donoghue1986low}
J.~F. Donoghue, E.~Golowich, and B.~Holstein, ``{Low-energy weak interactions
  of quarks},'' {\em Phys. Repts,} {\bfseries 131} (1986) 319--428.

\bibitem{barton1961notes}
G.~Barton, ``{Notes on the static parity non-conserving internucleon
  potential},'' {\em Il Nuovo Cimento} {\bfseries 19} (1961) 512--527.

\bibitem{glashow1970weak}
S.~L. Glashow, J.~Iliopoulos, and L.~Maiani, ``{Weak interactions with
  lepton-hadron symmetry},'' {\em Phys. Rev. D} {\bfseries 2} (1970) 1285.

\bibitem{kaplan1993analysis}
D.~B. Kaplan and M.~J. Savage, ``{An analysis of parity-violating pion-nucleon
  couplings},'' {\em Nucl. Phys. A} {\bfseries 556} (1993) 653--671.

\bibitem{Kai892}
N.~Kaiser and U.-G. Mei{\ss}ner, ``{The weak pion-nucleon vertex revisited},''
  {\em Nucl. Phys.} {\bfseries A489} (1988) 671--682.

\bibitem{Meissner:1998pu}
U.~G. Mei{\ss}ner and H.~Weigel, ``{The Parity violating pion nucleon coupling
  constant from a realistic three flavor Skyrme model},''
  \href{http://dx.doi.org/10.1016/S0370-2693(98)01569-X}{{\em Phys. Lett.}
  {\bfseries B447} (1999) 1--7},
\href{http://arxiv.org/abs/nucl-th/9807038}{{\ttfamily arXiv:nucl-th/9807038
  [nucl-th]}}.

\bibitem{Freeman:2012ry}
{\bfseries MILC} Collaboration, W.~Freeman and D.~Toussaint, ``{Intrinsic
  strangeness and charm of the nucleon using improved staggered fermions},''
  \href{http://dx.doi.org/10.1103/PhysRevD.88.054503}{{\em Phys. Rev.}
  {\bfseries D88} (2013) 054503},
\href{http://arxiv.org/abs/1204.3866}{{\ttfamily arXiv:1204.3866 [hep-lat]}}.

\bibitem{Oksuzian:2012rzb}
{\bfseries JLQCD} Collaboration, H.~Ohki, K.~Takeda, S.~Aoki, S.~Hashimoto,
  T.~Kaneko, H.~Matsufuru, J.~Noaki, and T.~Onogi, ``{Nucleon strange quark
  content from $N_f=2+1$ lattice QCD with exact chiral symmetry},''
  \href{http://dx.doi.org/10.1103/PhysRevD.87.034509}{{\em Phys. Rev.}
  {\bfseries D87} (2013) 034509},
\href{http://arxiv.org/abs/1208.4185}{{\ttfamily arXiv:1208.4185 [hep-lat]}}.

\bibitem{Junnarkar:2013ac}
P.~Junnarkar and A.~Walker-Loud, ``{Scalar strange content of the nucleon from
  lattice QCD},'' \href{http://dx.doi.org/10.1103/PhysRevD.87.114510}{{\em
  Phys. Rev.} {\bfseries D87} (2013) 114510},
\href{http://arxiv.org/abs/1301.1114}{{\ttfamily arXiv:1301.1114 [hep-lat]}}.

\bibitem{Gong:2013vja}
{\bfseries XQCD} Collaboration, M.~Gong {\em et~al.}, ``{Strangeness and
  charmness content of the nucleon from overlap fermions on 2+1-flavor
  domain-wall fermion configurations},''
  \href{http://dx.doi.org/10.1103/PhysRevD.88.014503}{{\em Phys. Rev.}
  {\bfseries D88} (2013) 014503},
\href{http://arxiv.org/abs/1304.1194}{{\ttfamily arXiv:1304.1194 [hep-ph]}}.

\bibitem{Dai:1991bx}
J.~Dai, M.~J. Savage, J.~Liu, and R.~P. Springer, ``{Low-energy effective
  Hamiltonian for $\Delta$ I = 1 nuclear parity violation and nucleonic
  strangeness},''
\href{http://dx.doi.org/10.1016/0370-2693(91)90108-3}{{\em Phys. Lett.}
  {\bfseries B271} (1991) 403--409}.

\bibitem{Buchalla:1995vs}
G.~Buchalla, A.~J. Buras, and M.~E. Lautenbacher, ``{Weak decays beyond leading
  logarithms},'' \href{http://dx.doi.org/10.1103/RevModPhys.68.1125}{{\em Rev.
  Mod. Phys.} {\bfseries 68} (1996) 1125--1144},
\href{http://arxiv.org/abs/hep-ph/9512380}{{\ttfamily arXiv:hep-ph/9512380
  [hep-ph]}}.

\bibitem{boyle2013emerging}
P.~A. Boyle, N.~H. Christ, N.~Garron, E.~J. Goode, T.~Janowski, {\em et~al.},
  ``{Emerging understanding of the $\Delta$ I= 1/2 rule from lattice QCD},''
  {\em Phys. Rev. Letts.} {\bfseries 110} no.~15, (2013) 152001.

\bibitem{Buras:2014maa}
A.~J. Buras, J.-M. Gerard, and W.~A. Bardeen, ``{Large $N$ approach to kaon
  decays and mixing 28 years later: $\Delta I = 1/2$ rule, $\hat B_K$ and
  $\Delta M_K$},'' \href{http://dx.doi.org/10.1140/epjc/s10052-014-2871-x}{{\em
  Eur. Phys. J.} {\bfseries C74} (2014) 2871},
\href{http://arxiv.org/abs/1401.1385}{{\ttfamily arXiv:1401.1385 [hep-ph]}}.

\bibitem{Sci5}
L.~Maiani and M.~Testa, ``{Final state interactions from Euclidean correlation
  functions},''
\href{http://dx.doi.org/10.1016/0370-2693(90)90695-3}{{\em Phys. Lett.}
  {\bfseries B245} (1990) 585--590}.

\bibitem{Sci6}
K.~Huang and C.~N. Yang, ``{Quantum-mechanical many-body problem with
  hard-sphere interaction},''
\href{http://dx.doi.org/10.1103/PhysRev.105.767}{{\em Phys. Rev.} {\bfseries
  105} (1957) 767--775}.

\bibitem{Sci7}
T.~D. Lee, K.~Huang, and C.~N. Yang, ``{Eigenvalues and eigenfunctions of a
  Bose system of hard spheres and Its low-temperature properties},''
\href{http://dx.doi.org/10.1103/PhysRev.106.1135}{{\em Phys. Rev.} {\bfseries
  106} (1957) 1135--1145}.

\bibitem{Sci8}
T.~T. Wu, ``{Ground state of a Bose system of hard spheres},''
\href{http://dx.doi.org/10.1103/PhysRev.115.1390}{{\em Phys. Rev.} {\bfseries
  115} (1959) 1390--1404}.

\bibitem{Sci9}
M.~Luscher, ``{Volume dependence of the energy spectrum in massive quantum
  field theories. 2. Scattering states},''
  \href{http://dx.doi.org/10.1007/BF01211097}{{\em Commun. Math. Phys.}
  {\bfseries 105} (1986) 153--188}.

\bibitem{Sci10}
M.~Luscher, ``{Two particle states on a torus and their relation to the
  scattering matrix},''
  \href{http://dx.doi.org/10.1016/0550-3213(91)90366-6}{{\em Nucl. Phys.}
  {\bfseries B354} (1991) 531--578}.

\bibitem{Sci36}
E.~Berkowitz, T.~Kurth, A.~Nicholson, B.~Joo, E.~Rinaldi, M.~Strother, P.~M.
  Vranas, and A.~Walker-Loud, ``{Two-nucleon higher partial-wave scattering
  from lattice QCD},''
  \href{http://dx.doi.org/10.1016/j.physletb.2016.12.024}{{\em Phys. Lett.}
  {\bfseries B765} (2017) 285--292},
\href{http://arxiv.org/abs/1508.00886}{{\ttfamily arXiv:1508.00886 [hep-lat]}}.

\bibitem{Tiburzi:2012xx}
B.~Tiburzi, ``{Isotensor hadronic parity violation},''
  \href{http://dx.doi.org/10.1103/PhysRevD.86.097501}{{\em Phys. Rev.}
  {\bfseries D86} (2012) 097501},
\href{http://arxiv.org/abs/1207.4996}{{\ttfamily arXiv:1207.4996 [hep-ph]}}.

\bibitem{Sci85}
D.~B. Kaplan, M.~J. Savage, and M.~B. Wise, ``{A new expansion for
  nucleon-nucleon interactions},''
  \href{http://dx.doi.org/10.1016/S0370-2693(98)00210-X}{{\em Phys. Lett.}
  {\bfseries B424} (1998) 390--396},
\href{http://arxiv.org/abs/nucl-th/9801034}{{\ttfamily arXiv:nucl-th/9801034
  [nucl-th]}}.

\bibitem{Sci86}
E.~Epelbaum, W.~Gloeckle, and U.-G. Mei{\ss}ner, ``{Nuclear forces from chiral
  Lagrangians using the method of unitary transformation. 1. Formalism},''
  \href{http://dx.doi.org/10.1016/S0375-9474(98)00220-6}{{\em Nucl. Phys.}
  {\bfseries A637} (1998) 107--134},
\href{http://arxiv.org/abs/nucl-th/9801064}{{\ttfamily arXiv:nucl-th/9801064
  [nucl-th]}}.

\bibitem{Sci87}
E.~Epelbaum, W.~Gloeckle, and U.-G. Mei{\ss}ner, ``{Nuclear forces from chiral
  Lagrangians using the method of unitary transformation. 2. The two nucleon
  system},'' \href{http://dx.doi.org/10.1016/S0375-9474(99)00821-0}{{\em Nucl.
  Phys.} {\bfseries A671} (2000) 295--331},
\href{http://arxiv.org/abs/nucl-th/9910064}{{\ttfamily arXiv:nucl-th/9910064
  [nucl-th]}}.

\bibitem{Sci88}
U.~van Kolck, ``{Effective field theory of nuclear forces},''
  \href{http://dx.doi.org/10.1016/S0146-6410(99)00097-6}{{\em Prog. Part. Nucl.
  Phys.} {\bfseries 43} (1999) 337--418},
\href{http://arxiv.org/abs/nucl-th/9902015}{{\ttfamily arXiv:nucl-th/9902015
  [nucl-th]}}.

\bibitem{Sci89}
S.~R. Beane, P.~F. Bedaque, M.~J. Savage, and U.~van Kolck, ``{Towards a
  perturbative theory of nuclear forces},''
  \href{http://dx.doi.org/10.1016/S0375-9474(01)01324-0}{{\em Nucl. Phys.}
  {\bfseries A700} (2002) 377--402},
\href{http://arxiv.org/abs/nucl-th/0104030}{{\ttfamily arXiv:nucl-th/0104030
  [nucl-th]}}.

\bibitem{Sci90}
E.~Epelbaum, H.-W. Hammer, and U.-G. Mei{\ss}ner, ``{Modern theory of nuclear
  forces},'' \href{http://dx.doi.org/10.1103/RevModPhys.81.1773}{{\em Rev. Mod.
  Phys.} {\bfseries 81} (2009) 1773--1825},
\href{http://arxiv.org/abs/0811.1338}{{\ttfamily arXiv:0811.1338 [nucl-th]}}.

\bibitem{Beane:2002ca}
S.~R. Beane and M.~J. Savage, ``{Hadronic parity violation on the lattice},''
  \href{http://dx.doi.org/10.1016/S0550-3213(02)00405-4}{{\em Nucl. Phys.}
  {\bfseries B636} (2002) 291--304},
\href{http://arxiv.org/abs/hep-lat/0203028}{{\ttfamily arXiv:hep-lat/0203028
  [hep-lat]}}.

\bibitem{Nollett}
L.~E. Marcucci, K.~M. Nollett, R.~Schiavilla, and R.~B. Wiringa, ``{Modern
  theories of low-energy astrophysical reactions},'' {\em Nucl. Phys.}
  {\bfseries A777} (2006) 111--136.

\bibitem{Viviani:2010qt}
M.~Viviani, R.~Schiavilla, L.~Girlanda, A.~Kievsky, and L.~E. Marcucci, ``{The
  parity-violating asymmetry in the $^3He(n,p)^3H$ reaction},''
  \href{http://dx.doi.org/10.1103/PhysRevC.82.044001}{{\em Phys. Rev.}
  {\bfseries C82} (2010) 044001},
\href{http://arxiv.org/abs/1007.2052}{{\ttfamily arXiv:1007.2052 [nucl-th]}}.

\bibitem{Desplanques:1986cq}
B.~Desplanques and J.~J. Benayoun, ``{Parity nonconserving effects in thermal
  neutron deuteron radiative capture},''
\href{http://dx.doi.org/10.1016/0375-9474(86)90194-6}{{\em Nucl. Phys.}
  {\bfseries A458} (1986) 689--708}.

\bibitem{Lobashov:1972fwg}
V.~M. Lobashov, D.~M. Kaminker, G.~I. Kharkevich, V.~A. Kniazkov, N.~A.
  Lozovoy, V.~A. Nazarenko, L.~F. Sayenko, L.~M. Smotritsky, and A.~I. Yegorov,
  ``{Parity non-conservation in radiative thermal neutron capture by
  protons},''
\href{http://dx.doi.org/10.1016/0375-9474(72)90759-2}{{\em Nucl. Phys.}
  {\bfseries A197} (1972) 241--258}.

\bibitem{Snow:2016zyq}
W.~M. Snow {\em et~al.}, ``{Status of theory and experiment in hadronic parity
  violation},''
\href{http://dx.doi.org/10.1142/S2010194516600028}{{\em Int. J. Mod. Phys.
  Conf. Ser.} {\bfseries 40} (2016) 1660002}.

\bibitem{Avishai:1985mu}
Y.~Avishai and P.~Grange, ``{Parity violation in threshold neutron proton
  scattering},''
\href{http://dx.doi.org/10.1088/0305-4616/10/12/001}{{\em J. Phys.} {\bfseries
  G10} (1984) L263--L270}.

\bibitem{Heckel1986}
B.~Heckel, ``{PNC and TNC rotations of the neutron spin},'' in {\em
  {Proceedings of the Workshop on the Investigation of Fundamental Interactions
  with Cold Neutrons}}, G.~L. Greene, ed., p.~90.
\newblock National Bureau of Standards, 1986.

\bibitem{Adelberger:1987zt}
E.~Adelberger, ``{Parity violation in nuclear systems, scattering and decay
  experiments},'' in {\em {Symposium/Workshop on Parity Violation in Hadronic
  Systems Vancouver, British Columbia, Canada, May 28-29, 1987}}, pp.~50--66.
\newblock
1987.
\newblock

\bibitem{Schiavilla:2004wn}
R.~Schiavilla, J.~Carlson, and M.~W. Paris, ``{Parity violating interaction
  effects in the np system},''
  \href{http://dx.doi.org/10.1103/PhysRevC.70.044007}{{\em Phys. Rev.}
  {\bfseries C70} (2004) 044007},
\href{http://arxiv.org/abs/nucl-th/0404082}{{\ttfamily arXiv:nucl-th/0404082
  [nucl-th]}}.

\bibitem{Griesshammer:2011md}
H.~W. Griesshammer, M.~R. Schindler, and R.~P. Springer, ``{Parity-violating
  neutron spin rotation in hydrogen and deuterium},''
  \href{http://dx.doi.org/10.1140/epja/i2012-12007-8}{{\em Eur. Phys. J.}
  {\bfseries A48} (2012) 7},
\href{http://arxiv.org/abs/1109.5667}{{\ttfamily arXiv:1109.5667 [nucl-th]}}.

\bibitem{Schiavilla:2008ic}
R.~Schiavilla, M.~Viviani, L.~Girlanda, A.~Kievsky, and L.~E. Marcucci,
  ``{Neutron spin rotation in n-polarized - d scattering},''
  \href{http://dx.doi.org/10.1103/PhysRevC.78.014002,
  10.1103/PhysRevC.83.029902}{{\em Phys. Rev.} {\bfseries C78} (2008) 014002},
  \href{http://arxiv.org/abs/0805.3599}{{\ttfamily arXiv:0805.3599 [nucl-th]}}.
[Erratum: Phys. Rev.C83,029902(2011)].

\bibitem{Dmitriev:1983mg}
V.~F. Dmitriev, V.~V. Flambaum, O.~P. Sushkov, and V.~B. Telitsin, ``{The
  parity violating rotation of the neutron spin in helium},''
\href{http://dx.doi.org/10.1016/0370-2693(83)91221-2}{{\em Phys. Lett.}
  {\bfseries B125} (1983) 1--4}.

\bibitem{Desplanques:1979st}
B.~Desplanques, J.~J. Benayoun, and C.~Gignoux, ``{Predictions of parity
  nonconserving effects in nucleon deuteron scattering},''
\href{http://dx.doi.org/10.1016/0375-9474(79)90580-3}{{\em Nucl. Phys.}
  {\bfseries A324} (1979) 221--233}.

\end{thebibliography}\endgroup

\end{document}